\documentclass[11pt,a4paper]{article}
\usepackage{jheppub}
\usepackage{amsmath,amssymb,epsfig}
\usepackage{upgreek}
\usepackage{color}
\usepackage{mathtools}
\usepackage[all,cmtip]{xy}

\renewcommand{\1}{\it 1}
\renewcommand{\2}{\it 2}
\newcommand{\3}{\it 3}

\newcommand{\C}{{\cal C}}
\newcommand{\G}{{\cal G}}
\renewcommand{\d}{{ d}}

\newcommand{\ls}{\ell_{\rm s}}

\newcommand{\ve}{\varepsilon}
\newcommand{\tr}{{\rm tr}}

\newcommand{\R}{{\mathbb R}}
\newcommand{\Z}{{\mathbb Z}}

\newcommand{\nn}{\nonumber}
\newcommand{\be}{\begin{equation}}
\newcommand{\ee}{\end{equation}} 
\newcommand{\bea}{\begin{eqnarray}}
\newcommand{\eea}{\end{eqnarray}}

\newcommand{\blue}{\color{blue}}

\title{Supergravity in Twelve Dimension}

\author{Kang-Sin Choi}
\affiliation{Scranton Honors Program, Ewha Womans University, Seoul 120-750, Korea}
\date{}
\emailAdd{kangsin@ewha.ac.kr}

\abstract{
We consider supergravity in twelve dimension, whose dimensional reduction yields eleven-dimensional, IIA, and IIB supergravities. This also provides the effective field theory of F-theory. We must take one direction as a compact circle, so that the Poincar\'e symmetry and the zero-mode field contents are identical to those of eleven-dimensional supergravity. We also have a tower of massive Kaluza--Klein states to be viewed as the wrapping modes of M2-branes. The twelfth dimension decompactifies only if other two directions are compactified on a torus, restoring different ten dimensional Poincar\'e symmetry of IIB supergravity, whose missing graviton is provided by components of the rank three tensor field. This condition prevents us from violating the condition on the maximal number of real supercharges, which should be thirty-two. The self-duality condition of the IIB  four-form fields is understood from twelve-dimensional Hodge duality.  In this framework $T$-duality is re-interpreted as taking different compactification routes.
}

\begin{document}

\maketitle


\section{Introduction}

This paper tries to tie up the loose ends of what have been known separately about duality relations of superstrings and supergraivities. We are particularly interested in the completion of IIB supergravity  \cite{Choi:2014vya}; First we consider a possibility to embed it into a certain higher dimensional supergravity. This turns out to require twelve dimension in total, in which the existence of supergravity is challenging. Also we want to concretely elucidate its $T$-duality to IIA supergravity. Although $T$-duality relation between type IIA and IIB supergravities compactified on circles is known \cite{TdualityII,Bergshoeff:1995as,Hassan,Jeon:2011sq}, we do not know which physics gives rise to such rules.

The formulation of superstring theories led us to discovery of a number of ten dimensional supergravity theories as the corresponding effective field theories. It has been soon realized that, between two possible supergravity theories of thirty-two supercharges, type IIA supergravity action can be obtained by dimensional reduction of an eleven-dimensional one, which is unique \cite{Cremmer:1978km}. The latter has less number of higher-dimensional multiplets and simpler form of interactions, at the price of introducing an extra dimension. Likewise, many features of realistic models are understood as geometric properties of extra dimensions \cite{KK}. For example, four-dimensional models describing our world are hoped to be obtained from a higher dimensional supergravity in a simpler form \cite{Beasley:2008dc,Heckman:2008qt,Donagi:2008ca,Tatar:2006dc,models,Abe:2008sx}. A natural question now is whether we can also obtain the other supergravity of type IIB by dimensional reduction of a higher dimensional theory.

It is well-known that type IIB superstring theory is more deeply understood as a dimensionally reduced one from a theory with two more dimensions, for the following reasons \cite{Vafa:1996xn,Morrison:1996na}. In the Einstein frame where the graviton kinetic term is canonical, the effective theory of IIB supergravity action is invariant under $SL(2,\R)$ symmetry, and kinetic terms for scalar fields take forms of moduli field of two extra dimensions. By string quantization the duality group reduces to $SL(2,\Z)$, which is the genuine symmetry of a torus \cite{Hull:1994ys}. Under this symmetry the complexified field $\tau$ of axion and dilation is identified as the complex structure of this torus, which depends on ten-dimensional spacetime. Now the problem is translated to find such twelve-dimensional theory concretly. Compactifications of internal dimensions for this twelve-dimensional theory are not possible unless we have a theory of gravity.  

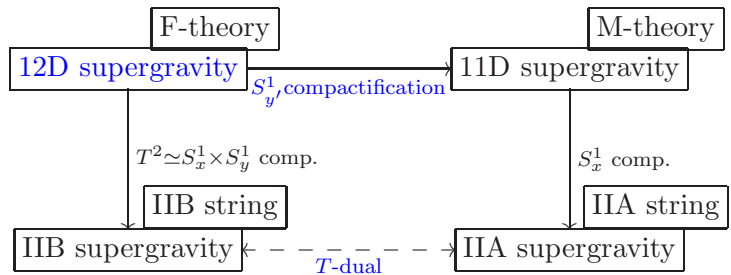
\begin{figure}[t]
$$
\xy
\xymatrixcolsep{9.3pc}
\xymatrixrowsep{4.3pc}
\xymatrix{
  *+[F]{\txt{F-theory }} &*+[F]{\txt{M-theory}} \\
*+[F]{\txt{IIB string}}  & *+[F]{\txt{IIA string}}
}
\POS-(14,5.3)
\xymatrixcolsep{6.4pc}
\xymatrixrowsep{4.4pc}
\xymatrix{
  *+[F]{\txt{\blue 12D supergravity}}  \ar[r]_{\blue S^1_{y'} \text{compactification}} \ar[d]^{T^2 \simeq S^1_x \times S^1_{y}\text{ comp.}} 
 &*+[F]{\txt{11D supergravity}} \ar[d]^{S^1_{x} \text{ comp.}} \\
*+[F]{\txt{IIB supergravity}} \ar@{<-->}[r]_{\text{\blue $T$-dual}}  & *+[F]{\txt{IIA supergravity}}
}
\endxy
$$
\caption{Relation among superstring and supergravity theories. The IIB theory can be interpreted to be obtained by compactifying the twelve-dimensional theory on torus. This $T$-duality holds between the two string theories if we remove the common directions. } \label{f:relations}
\end{figure}

Any attempt to construct twelve-dimensional supergravity with full Poincar\'e symmetry immediately faces obstacles. First, there is a classification on possible supermultiplets \cite{Nahm:1977tg}. 
The maximal number of real supercharges hence supersymmetry generators are thirty-two. In four dimensions, this brings the helicity $(-2)$ state of a graviton to a helicity $(+2)$ state, resulting in superpartner fields of gravitino and set of vector, scalar, spinorial fields. This set of supersymmetry generators and multiplets are best understood as dimensional reduction of simpler generators in higher dimensions, compactified on a higher dimensional torus. The maximal number of spacetime dimensions with Lorentz signature $(1,10)$ allowing this spinorial representation is eleven. Indeed, compactification of eleven dimensional supergravity theory on a circle gives ten-dimensional IIA supergravity with supercharge $(1,1)$ \cite{Cremmer:1978km}. The other ten-dimensional supergravity, IIB, has supercharge $(2,0)$. 
We cannot have supergravity with more supercharges since otherwise we must have a field with spin higher than two in four dimensions. No consistent interacting theory with spin higher two is known possible.  This is the reason why we cannot have twelve-dimensional supergravity with the full Poincar\'e symmetry. Noting that the Clifford algebra with Lorentz signature $(1,9)$ is isomorphic to that with $(2,10)$, one may try to construct thirty-two component spinor in the twelve dimension with two timelike directions. However this is not what we want, because the two extra dimensions of IIB theory should be compact thus space-like.

Then, what would be the form of twelve-dimensional supergravity, if there should exist a effective field theory of F-theory? This derivation of F-theory from M-theory hints us the meaning of the twelfth dimension \cite{Vafa:1996xn,Denef:2008wq}. It is well known that eleven dimensional supergravity is a field theory limit of M-theory \cite{Witten:1995ex,Horava:1995qa}. F-theory compactified on a torus $T^2 \simeq S^1_x \times S^1_{y'}$ is dual to M-theory compactified on the {\em same} torus in the zero size limit, thus we may expect the same relation also in the effective field theory limit. Low-energy effective actions are studied in this context \cite{Grimm:2010ks,Bonetti:2011mw,Grimm:2011fx,Malmendier:2014uka,Cvetic:2012xn,Martucci:2014ska,Junghans:2014zla,DelZotto:2014fia}.

To be more detailed, $T$-duality takes one of the torus cycle in the M-theory side, say $S^1_y$, to another dual circle $S^1_{y'}$ in the F-theory side. Since the radii of these circles are inversely related  $R_y = \ell_{\rm s}^2 / R_{y'}$ in the string length unit $\ell_{\rm s}$, in the zero size limit $R_y \to 0$ of one circle, the other circle $S^1_{y'}$ becomes decompactified, restoring different Lorentz symmetry in another ten dimension. Although we cannot maintain twelve-dimensional Lorentz symmetry fully, each ten- and eleven-dimensional theories are consitent in its own space. Still, however, F-theory possesses the torus, on which it is compactified to yield type IIB superstring. In other words, the torus is shared by both theories.

Therefore, it is natural to keep {\em both circle directions} together, although these two circles $S_y$ and $S_{y'}$ seems redundant \cite{Choi:2014vya}. Remarkably it turns out that there is no contradiction in the sense that we cannot see the twelve dimensions fully but only part of dimensions that we have known, and this shall be automatically taken into account by moduli fields. In this new framework, then, M-theory looks like a dimensional reduction of F-theory in the decompactifying torus limit, as schematically shown in Figure \ref{f:relations}. Following the program, we have presented the bosonic action in Ref. \cite{Choi:2014vya}. There have been many attempts to suggest possible terms in various contexts \cite{Ferrara:1996wv, Kar:1997cx, Donagi:2008kj}. In this paper, we complete the supergravity action up to linear order of fermions.

\begin{figure}[t]
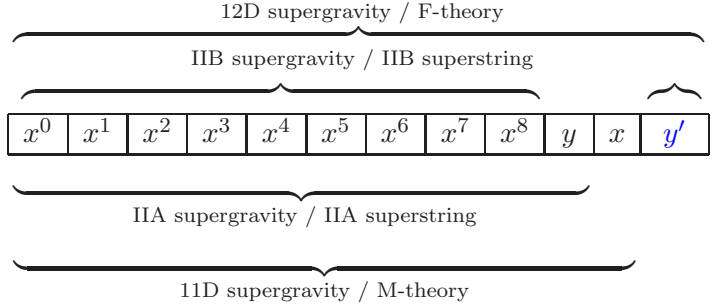

$$
\begin{array}{|c|c|c|c|c|c|c|c|c|c|c|c|}
\multicolumn{12}{c}{\overbrace{\rule{0.6\columnwidth}{0pt}}^{\text{ 12D supergravity / F-theory}}_{\text{ IIB supergravity / IIB superstring}}} \\
\multicolumn{9}{r}{\overbrace{\rule{0.45\columnwidth}{0pt}}} & \multicolumn{2}{c}{} & \multicolumn{1}{c}{\overbrace{\rule{0\columnwidth}{-0.2pt}}^{} } \\
\hline 
 ~x^0~& ~x^1~ & ~x^2~ & ~x^3~ & ~x^4~ & ~x^5~ & ~x^6~ & ~x^7~ & ~x^8~ & ~y~ & ~x~ & {\blue y'} \\
\hline
\multicolumn{10}{c}{\underbrace{\rule{0.5\columnwidth}{0pt}}_{\text{ IIA supergravity / IIA superstring}}} & \multicolumn{2}{c}{}\\
\multicolumn{11}{c}{\underbrace{\rule{0.54\columnwidth}{0pt}}_{\text{11D supergravity / M-theory}}} & \multicolumn{1}{c}{}\\
\end{array}
$$
\caption{Configuration of twelve-dimensional geometry and our convention of coordinate notations. The IIA supergravity lives in ten dimension having coordinates $(x^0,\dots,x^8,y)$, whereas another combination $(x^0,\dots,x^8,y')$ are spanned by the IIB supergravity, each of which has {\em different} ten-dimensional Poincar\'e symmetry. Only the time direction has the opposite Lorentz signature.}
\label{f:coordinates}
\end{figure}

Notation on coordinate indices and gamma matrices:
\begin{center} \small
\begin{tabular}{cccc} \hline 
dimension & general & local Lorentz & gamma matrices  \\ \hline
12 & $M,N,P,\dots$ & $A,B,C,\dots$ & $\Gamma^{A}$ \\
11 & $m,n,p,\dots$ & $a,b,c,\dots$ & $\Gamma^a$ \\
10 & $\upmu, \upnu, \uprho,\dots$ & $\upalpha,\upbeta,\upgamma,\dots$ & $\upgamma^\alpha$ \\
9 & $\mu,\nu,\rho,\dots$ & $\alpha, \beta,\gamma,\dots$ & $\gamma^\alpha$ \\ \hline
\end{tabular}
\end{center}
Sometimes we underline the curved coordinate when we need to distinguish.

Warning: To `derive' type IIB supergravity starting from a suggested twelve-dimensional action, we first work in the Einstein frame before Section \ref{sec:fermonic}, which enables us to `discover' string as one dimensional tensionful object. From Section \ref{sec:fermonic}, however, it is convenient to work in the string frame, whose details are given in the appendix.

\section{The twelve-dimensional bosinic action}

First we describe the bosonic part of twelve-dimensional supergravity. We should have one dimension compactified on a circle, so that at best we have eleven-dimensional Poincar\'e symmetry. We compare the {\em massless} degrees of freedom matches to those of eleven-dimensional supergravity.

\subsection{Twelve-dimensional action}

The bosonic degrees of freedom of eleven-dimensional supergravity are a graviton $G_{mn}$ and a rank-three antisymmetric tensor field $C_{mnp}$.
We lift eleven dimensional supergravity action to twelve-dimensional one. The three-form field is promoted to a four-form field as
\be \label{fourformpotential}
  C_{mnp}(x^m) \to {\C}_{mnpy'}(x^m,y').
\ee
having the dependence on an extra dimension, which we will call $y'$ hereafter. Also the graviton $G_{mn}$ is regarded as a part of the twelve-dimensional graviton. 
These can naturally appear if we have an orthogonal extra dimension with a metric
\be \label{twelvedmetric}
 ds^2 = \G_{MN} dx^M dx^N = G_{mn} dx^m dx^n + r^2 dy^{\prime 2},
\ee
where $M,N$ are twelve-dimensional, and $m,n$ are eleven-dimensional indices. Later we will come back to the issue of the off-diagonal component along $dx^m dy'$. 

We first suggest an action
\begin{equation} \label{twelveDaction}
 {2 \kappa_{12}^2} S =  \int d^{11}x dy' \sqrt{-\G} \left( {\cal R} -\frac12 |\G_{\it 5}|^2 \right)   + \frac{1}{6}\int {\cal C}_{\it 4} \wedge  G_{\it 4} \wedge G_{\it 4},
\end{equation}
and perform dimensional reductions shortly to consistently match known theories.
Here, $\cal R$ is the twelve-dimensional Ricci scalar made of the metric $\G_{MN}$ in (\ref{twelvedmetric}) and $\G$ is its determinant. The forms are defined as follows
\begin{align} 
{\cal C}_{\it4}   
 & \equiv C_{\it 3} \wedge r \d y' \nn \\
 & = \frac{1}{3!} r { C}_{mnp} \d x^m \wedge \d x^n \wedge \d x^p \wedge  \d y'  \label{C4def} \\
 & \equiv \frac{1}{4!} {\cal C}_{MNPQ} \d x^M \wedge \d x^N \wedge \d x^P \wedge  dx^Q |_{y'}, \nn \\
\G_{\it 5}& \equiv  G_{\it 4} \wedge r \d y' \equiv d C_{\3} \wedge r dy' = \d {\cal C}_{\it 4}  + C_{\3} \wedge dr \wedge dy'.
\label{G5}
\end{align}
Here the third line means we fix one of the component to be $y'$, for which we have four ways to do it.
In particular, the definition in the first line of (\ref{C4def}), written componentwise
\be \label{therefrom}
 C_{mnp} (x^m) r (x^m,y') = \C_{mnpy'} (x^m,y'),
\ee
suggests that $C_{\3}$ is a native eleven-dimensional field, which is related to the twelve-dimensional $\C_{\it 4}$ through the dependence on $y'$ in $r$. There have been attempts to treating them independent, which would violate the multiplet condition for the eleven-dimensional supergravity.

The norm and the wedge product are defined as
\begin{align}
|G_{\it p}|^2 &= \frac{1}{p!} \G^{M_1 N_1} \G^{M_2 N_2} \cdots \G^{M_p N_p} G_{M_1 M_2 \dots M_p} G_{N_1 N_2  \dots N_p}, \\
{\C}_{\it 4} \wedge G_{\it 4} \wedge G_{\it 4} &= \frac{1}{3! 4! 4!} \epsilon_{m n p y' m_5 m_6 \dots m_{12}} \C_{m n p y'} G_{m_5 \dots m_8} G_{m_9 \dots m_{12}} dx^{m_1}  dx^{m_2}  \cdots dx^{m_{12}}.
\end{align}
Here $\epsilon_{m_1 \dots m_{12}}$ is the totally antisymmetric Levi-Civita symbol. Note that $\G_{\it 5}$ is not the exterior derivative of $\C_{\it 4}$, which is usually the case of dimensionally reduced theories. We will define $\kappa_{12}$ soon. 
It is important to note that the indices assume only eleven-dimensional coordinates and we have incomplete components for $\C_{\it 4}$. Therefore the action (\ref{twelveDaction}) has at best {\em eleven-}dimensional Poincar\'e symmetry. Nevertheless this form is useful, since we may recover ten-dimensional symmetry in which we {\em include} the $y'$-direction. 

Finally, there is an extra term from one-loop contribution
\be
 S_{\rm one} = - \frac{2 \pi \ell^7}{2 \kappa_{12}^2 }  \int \C_{\it 4}\wedge I_8,
\ee
where $I_8$ is a polynomial only dependent on Ricci tensors, which can be found in \cite{Duff:1995wd} 
$$ I_8 = \frac{1}{(2\pi)^4}\left( -\frac{1}{768}(\tr R^2)^2 + \frac{1}{192} \tr R^4 \right). $$

The third term in (\ref{twelveDaction}), of a Chern--Simons type, has a property 
\be \label{CSterm}
 {12 \kappa_{12}^2} S_{\rm CS}= \int \C_{\it 4} \wedge G_{\it 4}  \wedge G_{\it 4}  = \int C_{\it 3} \wedge r dy'  \wedge G_4 \wedge G_{\it 4} = - \int \G_{\it 5} \wedge C_{\3} \wedge G_{\it 4},
\ee
which looks like integration by parts in the end, although what we actually did is to just follow the definition (\ref{G5}) and changed the order of the wedge products. With this, varying $C_{\it 3}$, we obtain the equation of motion for $\C_{\it 4}$ 
\begin{equation} \label{EOM}
 \d {* \G_{\it 5}} = - \frac12 d(C_{\3} \wedge G_{\it 4})= - \frac12 G_{\it 4} \wedge G_{\it 4},
\end{equation}
supplemented by the Bianchi identity
$$ 
\d\G_{\it 5} = G_{\it 4} \wedge dr \wedge dy'.
$$
This will not affect the Bianchi identity of $G_{\it 4}$
$$d G_{\it 4} = d^2 C_{\3}=  0,$$
which directly follows from the definition. 

Exchanging the role of the two, we also have a dual field strength $G_{\it 7}$
\begin{equation} \label{duality}
 G_{\it 7} \equiv * \G_{\it 5}  \equiv \d {C}_{\it 6} - \frac12 C_{\it 3} \wedge G_{\it 4},
\end{equation}
which defines a six-form ${C}_{\it 6}$. Componentwise, we have
\be \label{dualcomp}
 G_{m_1 \dots m_7} = \frac{1}{4!} \sqrt{-\G} \epsilon_{m_1 \dots m_{11} y'}  \G^{m_8 n_8} \cdots \G^{m_{11} n_{11}} \G^{y' y'} {\cal G}_{n_8 n_9 n_{10} n_{11} y'}  ,
\ee
where the indices are contracted by twelve-dimensional metric (\ref{themetric}). 
Since the totally antisymmetric tensor in the twelve-dimensional in (\ref{dualcomp}) takes one index on $y'$, $C_{\it 6}$ cannot have an index on $y'$.\footnote{In this paper, the curly letters have dependence on the $y'$-direction while the printed letters do not.} Although similar Hodge dual operation is possible in eleven-dimensional theory as well, this constraint is unique if we embed it in twelve dimension.

\subsection{Reduction to eleven-dimension}

We compactfy the extra dimension by identifying the coordinate
\be \label{elldefinition}
 y' \sim y' + 2 \pi \ell,
\ee
where $\ell$ is a length unit. Neverthless $\ell$ does not appear in the metric because of general coordinate invariance that always reduces the length scale in the $y'$-direction.

The Klein decomposition gives rise to infinite tower of fields 
\be \label{KKtower} \begin{split}
 \G_{mn}(x^m,y') &= \sum_{k=-\infty}^{\infty} \G_{mn}^{[k]} (x^m) e^{i k y' / \ell}, \\
 \C_{mnpy'}(x^m,y') &= \sum_{k=-\infty}^{\infty} \C_{mnpy'}^{[k]} (x^m) e^{i k y' / \ell}, \\
  \G_{mnpqy'}(x^m,y') &= \sum_{k=-\infty}^{\infty} \G_{mnpqy'}^{[k]} (x^m) e^{i k y' / \ell}, \\
    r(x^m,y') &= \sum_{k=-\infty}^{\infty} r^{[k]} (x^m) e^{i k y' / \ell}.
\end{split}
\ee
Since we have the invariant `geodesic' radius $\ell r$ as eigenvalue of momentum operator in the eleven dimensions, each mode has KK mass
\be \label{KKmasses}
 M_{k}^2 = \frac{k^2}{\ell^2 \langle r \rangle^2}.
\ee
In the effective eleven-dimensional action (\ref{11Daction}), we keep the zero mode in the action after renaming
$$ 
 G_{mn} (x^m) \equiv \G^{[0]}_{mn} (x^m), \quad C_{mnp} (x^m) \equiv (r^{[0]})^{-1} \C^{[0]}_{mnpy'} (x^m) , \quad G_{mnpq} (x^m) \equiv (r^{[0]})^{-1} \G^{[0]}_{mnpqy'} (x^m), 
 $$
and truncate the other massive modes $k \ne 0$ of masses (\ref{KKmasses}) considering small $r$ limit.\footnote{As long as there is no confusion, we use the same letters $C_{\it 3},G_{\it 4}$ for the twelve dimensional fields $C_{\it 3}(x^M),G_{\it 4}(x^M)$ and their eleven-dimensional zero modes $C_{\it 3}(x^m),G_{\it 4}(x^m)$.} This is consistent with the reliation (\ref{therefrom}). These modes will be interpreted as wrapping mode of M2-branes later. 

It should be noted that although the zero mode of $\C_{\it 4}$ and $\G_{\it 5}$ consistently give the eleven-dimensional field and its field strength $C_{\3}$ and $G_{\it 4}$, it cannot be said that we could derive the eleven-dimensional fields by reduction. In fact we have defined the former in terms of the latter by multiplying $r(x^m,y')$. Therefore at best we can check that we have the correct field degrees of freedom. This is because this dynamics is due to the behavior of the M2-branes and cannot be completely explained by this field theory limit (for example in string picture, this is related to winding string in the $T$-dual theory).

Now consider the action. The first two terms in (\ref{twelveDaction}) give
$$
\frac{1} {2\kappa_{12}^2}  \int_{12\rm D} d^{12} x \sqrt{-\G} \left( {\cal R} -\frac12 |\G_{\it 5}|^2 \right) =  \frac{1} {2\kappa_{12}^2} \int_{S^1} dy' \int_{11\rm D} d^{11}x \sqrt{-G}  r\left( R -\frac12 | G_{\it 4}|^2 \right) ,
$$
with a total derivative of a function of $r$, which can be removed by an appropriate boundary condition. Further compactification and consideration in lower dimensional theory shall later require the condition for $r$. After settling down everything we may consider the vacuum expectation value of $r$. With this, we {\em define} $\kappa_{12}$ in terms of $\kappa_{11}$ in ({\ref{11Daction}),
\begin{equation} \label{kappa12}
 \frac{2 \pi \ell \langle r \rangle } {2\kappa_{12}^2} = \frac{1}{2 \kappa_{11}^2}.
\end{equation} 
In this process, we may learn that the definition of $\G_{\it 5}$ in (\ref{G5}) is the only possibility to give the desired kinetic term for $C_{\it 3}$, while the other seemingly plausible choice $\G_{\it 5} = d \C_{\it 4}$ cannot.
Note that the Chern--Simons action (\ref{CSterm}) also gives rise exactly the same factor as in (\ref{kappa12}).
Therefore, we have obtained the bosonic part of the eleven-dimensional supergravity action
\begin{equation} \label{11Daction}
{2 \kappa_{11}^2} S =  \int d^{11}x \sqrt{-G} \left( R -\frac12 | G_{\it 4}|^2 \right)  - \frac16 \int { C}_{\it 3} \wedge G_{\it 4} \wedge G_{\it 4} ,
\end{equation}
The action (\ref{11Daction}) is also regarded as a low-energy, field-theoretical, effective action of M-theory, in the limit the that the fundamental object of M2-brane becomes pointlike. Therefore, we should have more irrelevant interaction operators at higher energies. Some of them are provided by Kaluza--Klein (KK) towers of fields in (\ref{KKtower}).

We have a scalar field $r$, which may raise two questions. First, this would exceed the desired degree of freedom of eleven-dimensional supergravity, failing to form the supersymmetric multiplet. We will see later that this will be related to the components of the graviton. Also, it is questionable whether the action is consistent in eleven and twelve dimensions in the sense that the resulting Einstein equation is solvable. It is known that simply turning off the scalar in the dimensional reduction of pure gravity is inconsistent \cite{Duff:1986hr}. In particular what concerns us is the ${y'y'}$ component of the twelve-dimensional Einstein equation. In our case, we are turning off the vector field while keeping the scalar field, and this is even possible from pure gravity, as a special case of Brans--Dicke theory \cite{Overduin:1998pn}. Since we are not dealing with pure gravity, but we have source terms through $\C_{\it 4}$ from the third and the last terms on the right-hand side (RHS) of (\ref{twelveDaction}), it seems always possible to satisfy the equation.

\section{Reduction to the IIB bosonic action}

By compactification on a circle on $y'$-direction, we have obtained eleven-dimensional supergravity. Further toroidal compactification gives supergravity actions of thirty-two supercharges.

\subsection{Nine dimensional geometry}

We compactfy further two dimensions on a torus of side length $L$ in the unit $\ell$ used (\ref{elldefinition}), with a complex structure 
\be \label{cs}
 \tau = \tau_1 + i \tau _2
\ee
This is done by identifying two coordinates as
$$
 x \sim x + 2 \pi \ell  \sim x + \tau y+ 2 \pi \tau \ell.
$$
The observed lengths are always given in combination of the metric
\begin{equation} \label{themetric}
 \d s^2 =  L^2 \left(\d x +  \tau_1 \d y +(a_{\mu} - \tau_1 b_{ \mu}) \d x^\mu  \right)^2 
                    +  L^2 \tau_2^2 \left( \d y -  b_{\mu } \d x^\mu\right)^2 
                       +r^2 \d y^{\prime 2} + g_{\mu \nu}' dx^\mu dx^\nu.
\end{equation}
It is the most general metric leaving the $y'$ direction orthogonal. Here, the metric components $a_{\mu}$ and $b_{\mu}$ are now Lorentz vectors in nine dimension, promoting the $S^1$ isometries of the $x$ and $y$ directions, respectively, to $U(1)$ gauge symmetries. 

In the classical theory, this torus has $SL(2,\R)$ symmetry due to diffeomorphism invariance of the covariance of general relativity. This is the reason why the radius do not appear in the metric (\ref{themetric}).
If a quantum theory of gravity breaks the scale invariance and set the fundamental length $\ell$, then we would have just $SL(2,\Z)$, generated by $\tau \to \tau +1$ and $\tau \to -1 /\tau$. This is what happens in Type IIB string theory or F-theory \cite{Hull:1994ys}. String quantization reduces the continuous internal symmetry to discrete. Now the origin of $SL(2,\R)$ symmetry of type IIB theory is identified as the symmetry of the torus. $\tau$ is also interpreted as the complex structure and transforms under this modular group. The pair $A_{\mu \nu},B_{\mu \nu}$ originates from the dimensional reduction of $\C_{\it 4}$ one of whose coordinates are respectively fixed to be $y$ and $x$.

To facilitate calculation, we introduce zw\"olfbeins \cite{Becker:2007zj}. By rewriting the metric (\ref{themetric})
\begin{equation} \G_{MN} = L^2
\begin{pmatrix}
  L^{-2} g_{\mu \nu}' + R_\mu R_\nu +  I_\mu I_\nu   & \tau_1 R_\mu + \tau_2 I_\mu  &  R_\mu & 0 \\
 \tau_1 R_\nu + \tau_2 I_\nu  & \tau_1^2 + \tau_2^2 & \tau_1 & 0 \\
  R_\nu &  \tau_1 & 1 & 0 \\
 0  &  0 & 0 & L^{-2} r^2  
\end{pmatrix},
\end{equation}
where $R_\mu = a_{\mu } - \tau_1b_{\mu }, I_\mu = -\tau_2 b_{\mu}$ and the bases of the block submatrices are understood, we obtain
\be
 e^A_M =  
 \begin{pmatrix}
 e^\alpha_{\mu}  & 0 & 0 & 0\\ 
 - L I_\mu  & L \tau_2 & 0 & 0 \\
 L R_\mu  & L \tau_1 & L   & 0 \\
 0 & 0 & 0 & r 
\end{pmatrix}, \quad G_{MN} = e_M^A e_N^B \eta_{AB},
\ee 
where $\eta_{AB}={\rm diag\, }(-1,1,1,\dots, 1)$ is a rank twelve $c$-number matrix. It is useful to define its inverse,
\begin{equation} \label{Invzwoelfbein}
 E^M_A = 
\begin{pmatrix}
 E^\mu_{\alpha}  & 0 & 0 &0 \\ 
 b_{\alpha} & (L \tau_2)^{-1} & 0 &0  \\
 - a_{\alpha} & -\tau_1(L \tau_2)^{-1} & L^{-1} & 0 \\   
 0 & 0 & 0 & r^{-1} 
\end{pmatrix},
\quad e_M^A E^M_B = \delta^A_{B}.
\end{equation}
For applying the metric, it is convenient to work in local Lorentz frame by using zw\"olfbeins,
\be \label{fieldrecuction}
 \begin{split}
 \C_{ABCD} &= \G_{MNPQ} E^M_A E^N_B E^P_C E^Q_D, \\ 
 \G_{EABCD} &= \G_{LMNPQ} E^L_E E^M_A E^N_B E^P_C E^Q_D . 
 \end{split}
\ee
Then the fields in (\ref{fieldrecuction}) are invariant under the gauge symmetries originating from the isometries. 
In fact, due to eleven-dimensional structure and the block-diagonal form of the zw\"olfbein (\ref{Invzwoelfbein}), we will always consider the case with $E^Q_D = E^{y'}_{y'} = r^{-1}$.
After the reduction, in the low dimension, we can recover the curved index structure by multiplying the vielbeins.

\subsection{Field reduction and decompactification}

\begin{table}
\begin{center}
 \begin{tabular}{cccc}
 \hline \hline
10D field & type & (9+1)D components & 12D components\\ \hline
$A_{\it 1}$ & RR & $\{ A_{\mu}, A_{y}\}$ & $\{ a_\mu,\tau_1\}$ \\
$A_{\it 3}$ & RR & $\{A_{\mu \nu \rho}, A_{\mu \nu y}\}$ & $ \{r^{-1} \C_{\mu \nu \rho y' },r^{-1} \C_{\mu \nu y y'}  \}$\\
$B_{\it 2}$ & NSNS & $ \{B_{\mu \nu },B_{\mu y} \}$ & $\{r^{-1} \C_{\mu \nu x  y'}, r^{-1} {\C}_{ \mu x y y' } \}$ \\
$b_{\it 1}$ & KK & $b_{\mu} $ & $b_\mu$ \\
\hline
 $A_{\it 4}$ & RR & $ A_{\mu \nu \rho y'}  $ & $r^{-1}{\C}_{\mu \nu \rho y' } $ \\
   $A_{\it 2}$ &RR &$\{A_{\mu \nu}, A_{\mu y'} = -A_{y' \mu} \}$ & $\{r^{-1} {\C}_{\mu \nu y y'},a_{ \mu}\}$ \\
 $A_{\it 0}$ & RR & $ A$ & $\tau_1$ \\
 $B_{\it 2}$ &  NSNS & $ \{-B_{\mu \nu},B_{\mu y'} = - B_{y' \mu}  \}$ & $\{r^{-1}{\C}_{\mu \nu x y' },b_{\mu}\}$ \\
 $K_{\it 1}$ & KK  & $K_{\mu} $ & $r^{-1} {\C}_{\mu x y y' }$ \\
\hline 
\end{tabular} 
\end{center}
\caption{Identification of ten-dimensional fields as collections of nine-dimensional fields. The upper and lower subtables respectively correspond to  IIA and IIB supergravity. Indices are nine-directional and $y'$ denotes the twelfth direction. Componentwise $\C_{mnpy'}=r C_{mnp}$ as in (\ref{C4def}). After decompactifying $y'$ or $y$ directions ten-dimensional Lorentz covariance is recovered. Also their magnetic dual fields follows from Hodge duality in twelve-dimension \cite{Choi:2014vya}.} \label{t:fields}
\end{table} 

The nine-dimensional fields are obtained by reduction of the four-form field and the metric tensor in twelve dimension, as in Table \ref{t:fields}. The normalization is more natural if we start from eleven-dimensional supergravity, but the tensor structure of the four-form field $A_{\it 4}$ is more natural in twelve-dimensions.

First, a rank two form field in the Neveu--Schwarz--Neveu--Schwarz (NSNS) comes from the following zero modes
\be 
\begin{split}
\C_{\mu \nu x y'} &\equiv - r B_{\mu \nu}, \quad  \G_{\mu \nu \rho x y'}  \equiv  -r  H_{\mu \nu \rho} = -3 r \partial_{[\mu} B_{\nu \rho]},
\end{split}
\ee
We are using the standard antisymmetric tensor notation by square brackets \cite{MTW}. We have to take $L \to 0$ to decouple the degrees of freedom depending on $x$ and $y$.

For example, consider
\be
\begin{split}
 \G_{\alpha \beta \gamma x  y'} & = \G_{[\mu \nu \rho] x y'} E^{\mu}_{\alpha} E^\nu_\beta E^\rho_{\gamma} E^x_{x} E^{y'}_{y'} + 3\G_{[\mu \nu  y] x y'} E^{\mu}_{\alpha} E^\nu_\beta E^y_{\gamma } E^x_{x} E^{y'}_{y'}  \\
  & = L^{-1} ( H_{\alpha \beta \gamma} +3 b_{[\alpha} H_{\beta \gamma]}),  \label{H3} 
\end{split}
\ee
where 
\be
 H_{\alpha \beta}  \equiv 2 \partial_{[\alpha} K_{\beta]}.
 \ee
The relation in the parenthesis in (\ref{H3}) can be recasted as
\begin{equation} \label{recovery} \begin{split}
 H_{\alpha \beta \gamma} +3 b_{[\alpha} H_{\beta \gamma]} &= H_{\alpha \beta \gamma} + 6 b_{[\alpha} \partial_{\beta} K_{\gamma]} \\
 & = H_{\alpha \beta \gamma} + 6 K_{[\alpha} \partial_\beta b_{\gamma] } + 6 \partial_{[\alpha} (K_{\beta} b_{\gamma]}) \\
 & = H_{\alpha \beta \gamma} + 3 K_{[\alpha} h_{\beta \gamma] }
\end{split}
\end{equation}
In the last line, the total derivative is a gauge symmetry of $H_{\alpha \beta \gamma}$. Now, the form (\ref{recovery}) can be regarded as arisen one from compactification of IIB supergravity
\be \label{oxidation}
H^{(10)}_{\mu \nu \rho}=H_{\mu \nu \rho}, \quad H^{(10)}_{\mu \nu y'} = h_{\mu \nu} \equiv  2 \partial_{[\mu} b_{\nu]},  \quad 
 b_{\alpha} = r B^{(10)}_{\alpha y'},
\ee
using the metric
\begin{equation} \label{recoveredmetric}
\begin{split}
 \d s^2 &=  L^2 \left(\d x +  \tau_1 \d y \right)^2  +  L^2 \tau_2^2  \d y^2 
                       +r^2 (\d y^{\prime} + K_\mu dx^{\mu})^2 + g_{\mu \nu}' dx^{\mu} dx^{\nu} \\
             & \Rightarrow L^2 \left(\d x +  \tau_1 \d y \right)^2  +  L^2 \tau_2^2  \d y^2 + G_{\upmu \upnu}' dx^{\upmu} dx^{\upnu}     .
\end{split}
\end{equation}
The justification is highly nontrivial, but we observe the following footprints whenever we perform dimensional reduction as in (\ref{themetric}).
\begin{enumerate}
\item A $U(1)$ gauge boson, usually called KK gauge boson, becomes a component of graviton in the extra dimension. This means, the ten-dimensional metric is re-arranged to include the vector field $K_\mu =r^{-1} {\cal C}_{\mu xyy'} $, gauging the isometry in the $y'$ direction. 
\item Appropriate additional terms for every reduced field coupled with KK gauge boson, by which the ten-dimensional fields are fully covariant. This condition {\em dictates} us to (from) which direction the (un)compactication takes place. 
\item The tower of the KK states for every reduced field.  In this case, we need towers of $g^{\prime k}_{\mu \nu}(x^{\mu}), B^k_{\mu \nu}(x^{\mu}), b^k_{\mu}(x^{\mu}), K^k_{\mu}(x^{\mu})$ with exactly the same mass squared $k^2 (\langle r \rangle \ell)^{-2}$ to each other. They exist because all of them come from the reduction of twelve-dimensional fields (\ref{KKtower}) with the physical $y'$-direction. 
\end{enumerate}
What is special in this case of dimensional reduction of twelve-dimensional supergravity is the following. For the first condition, the KK gauge boson $K_\mu$ did not come from dimensional reduction of higher dimensional graviton $a_\mu$ or $b_\mu$, but from antisymmetric tensor field ${\cal C}_{\mu xyy'}$. For this, we may interpret that the metric tensor itself is {\em not observable}, but indirectly measurable only by interaction with other fields. This is related to the second condition: Besides the original $y$-direction from which we got KK gauge boson $b_{\mu}$, we have shown in (\ref{recovery}) that the nine dimensional theory has another {\em another} $y'$-direction to which we can uncompactify with $K_{\mu}$. We will see later that these two decompactifications are exclusive.  For the third condition, the decompactification does not happen if we start from ten-dimensional IIA or the eleven-dimensional supergravities, since the KK states are missing. However they provide at best indirect evidence of the presence of the limit of uncompact $y'$-direction recovering ten-dimensional Lorentz symmetry. The only plausible option at the moment seems to be the formulation of type IIB string theory in ten dimensions, independent of the formulation of M-theory.

Now we perform dimensional reduction for Ramond--Ramond (RR) four-form as
\begin{equation}  
A_{\mu \nu \rho y'} \equiv r^{-1} {\cal C}_{\mu \nu \rho y'}, \quad F_{\mu \nu \rho \sigma y'} = { 4 \partial_{[\mu}  A_{\nu \rho \sigma] y'} }\equiv r^{-1} {\G}_{\mu \nu \rho \sigma y'}
\end{equation}
but only a part of it: one of whose index is fixed in the $y'$ direction.
Applying the same procedure, we have
\begin{align}
 \G_{\alpha \beta \gamma\delta y'} & = \G_{\mu \nu \rho \sigma y'} E^{\mu}_{\alpha} E^\nu_\beta E^\rho_\gamma E^\sigma_{\delta} E^{y'}_{y'} +4\G_{\mu \nu \rho x y'} E^{\mu}_{\alpha} E^\nu_\beta E^\rho_\gamma E^x_{\delta} E^{y'}_{y'} + 4\G_{\mu \nu \rho y y'} E^{\mu}_{\alpha} E^\nu_\beta E^\rho_\gamma E^y_{\delta} E^{y'}_{y'}  \nn \\
  & \quad +12 \G_{\mu \nu x y y'} E^{\mu}_{\alpha} E^\nu_\beta E^x_\gamma E^y_{\delta} E^{y'}_{y'}  \nn \\
& = F_{[\alpha \beta \gamma \delta] y'} + 4 a_{ [\alpha} H_{\beta \gamma \delta]} - 4 b_{ [\alpha} F_{\beta \gamma \delta]} 
+ 12 a_{[\alpha} b_\beta H_{\gamma \delta]}   \nn \\
& = F_{[\alpha \beta \gamma \delta] y'} + 4 a_{ [\alpha} H_{\beta \gamma \delta]} -4 b_{ [\alpha} F_{\beta \gamma \delta]} + 12a_{[\alpha} K_{\beta} h_{\gamma \delta]} -12 b_{ [\alpha} K_{\beta} f_{\gamma \delta]} \label{G5exp}  \\
&=  F_{[\alpha \beta \gamma \delta]y'} + 2 a_{[\alpha} ( H_{\beta \gamma \delta]} +3 K_{\beta} h_{\gamma \delta]}) + 3 f_{ [\alpha \beta} (B_{\gamma \delta]} +2  K_{\gamma} b_{\delta] }) \nn \\
&\quad - 2 b_{ [\alpha} ( F_{\beta \gamma \delta]} +3 K_{\beta} f_{\gamma \delta]}) - 3 h_{ [\alpha \beta} (A_{\gamma \delta]} +2  K_{\gamma} a_{\delta] }), \nn 
\end{align}
where we defined
$$ 
 f_{\alpha \beta} \equiv  2 \partial_{[\alpha}  a_{\beta]}, \quad h_{\alpha \beta} \equiv 2 \partial_{[\alpha} b_{\beta]}.
$$
The result shows that we have the same dimensionally reduced structure as in (\ref{recovery}), satisfying the above three conditions.
Note that the seemingly inverted relations (\ref{fielddecomp}) are because they are relations of local flat space. Rewriting the metric into the inverse {\em zehnbein}, obtained from the ten-dimensional metric (\ref{recoveredmetric})
\be
 E^{\upmu}_{\upalpha} = \begin{pmatrix} E^{\mu}_\alpha & 0 \\ K _{\alpha} & r^{-1} \end{pmatrix},
\ee
we can easily calculate
\be 
\begin{split} \label{indexrule}
 F^{ (10)}_{\alpha \beta \gamma \delta y'} 
 &= E^{\mu}_\alpha E^\nu_\beta E^\rho_\gamma E^{\sigma}_\delta E^{y'}_{y'} F^{(10)}_{\mu \nu \rho \sigma y'} \\
 &= r^{-1} F_{\alpha \beta \gamma \delta y'},
\end{split}
\ee
where the field in the last line is nine-dimensional. Likewise we obtain well-known Buscher relations between antisymmetric tensor fields \cite{Bu}
\be \label{fielddecomp}
 a_{\alpha} = r  A^{(10)}_{\alpha y'}, \quad A_{\alpha \beta} + 2 a_{[\alpha} K_{\beta]} = A^{(10)}_{\alpha \beta}, \quad  B_{\alpha \beta} + 2 b_{[\alpha} K_{\beta]} = B^{(10)}_{\alpha \beta},
\ee
and similar for their field strengths. 
Therefore, rewriting (\ref{G5exp}) we have
\begin{align}
r( F^{(10)}_{[\alpha \beta \gamma \delta] y'} &
+ 2A^{(10)}_{y' [\alpha} H^{(10)}_{\beta \gamma \delta]} 
 - 3 A^{(10)}_{[\alpha  \beta} H^{(10)}_{\gamma \delta]y'}
 - 2 B^{(10)}_{y' [\alpha} F^{(10)}_{ \beta \gamma \delta]} 
  +3 B^{(10)}_{[\alpha \beta}  F^{(10)}_{ \gamma \delta]y'} )\nn \\
&=r \left(F_{\it 5}^{(10)}-  \frac12 A_{\2}^{(10)}\wedge H_{\3}^{(10)}  + \frac12  B_{\2}^{(10)} \wedge F_{\3}^{(10)} \right)_{[\alpha \beta \gamma \delta] y'}\label{G5exp10D} \\
& \equiv r \tilde F^{(10)}_{[\alpha \beta \gamma \delta]y'}, \nn
\end{align}
using the metric (\ref{recoveredmetric}), recovering the total antisymmetric structure. The complete antisymmetric structure comes from total antisymmetrization of $\C_{\it 4}$ and $\G_{\it 5}$, which is only possible in twelve-dimensional lift. 
For this special four-form field, we need the other component to recover ten-dimensional Poincar\'e symmetry. Since $\C_{\it 4}$ can at best give rank four field in ten dimensions, we consider its magnetic dual field in the next subsection. Other components are dimensionally reduced in similar fashion, shown in the appendix.

\subsection{Self-duality of the IIB four-form field}

We have seen that the Hodge duality of $\G_{\it 5}$ in (\ref{duality}) defined the six-form $C_{\it 6}$. In this, total antisymmetric tensor in Hodge duality relation in twelve-dimension prevented the $C_{\it 6}$ from taking an index on $y'$. With three compactified directions along $x,y,y'$, the four-form field $A_{\it 4}$ is obtained only by fixing two indices of the $C_{\it 6}$ to be on $x$ and $y$,
\begin{equation} \label{C4fromC6}
 {C}_{\mu \nu \rho \sigma x y} \equiv  A_{\mu \nu \rho \sigma}, \quad  F_{\mu \nu \rho \sigma \tau} \equiv 5 \partial_{[\mu} A_{\nu \rho \sigma \tau]}.
 \end{equation}
Again we emphasize that this $C_{\it 6}$ is {\em defined} through the twelve-dimensional Hodge duality (\ref{duality}), or in components (\ref{dualcomp}).
To perform dimensional reduction, we go to local Lorentz frame in which $\sqrt{-\G} =1$ and $\G_{AB}=\eta_{AB}$. Going to nine dimension and performing decompactification to ten dimensions, the right-hand side of (\ref{dualcomp}) gives what we have just computed in (\ref{G5exp10D})\footnote{Note that there is cancellation between $r$ in the definition of $r F_{\mu \nu \rho \sigma y'}$ and $E^{y'}_{y'}$ in the nine-dimensional relation, there is another $r$ factor as in (\ref{G5exp10D}), in going to ten dimension.}
\be
 \begin{split}
  \frac{1}{5!} \epsilon_{\alpha_1 \alpha_2 \dots \alpha_{9} x y y'}  r  \tilde F_{\alpha_6 \alpha_7 \alpha_8 \alpha_9  y'}^{(10)} = r {*_{10} }\tilde F^{(10)}_{\alpha_1 \alpha_2 \alpha_3 \alpha_4 \alpha_5} .
\end{split}
\ee
Taking into account the factors $E^x_x=L^{-1}, E^y_y=(L \tau_2)^{-1},$ the left-hand side (LHS) of (\ref{dualcomp}) becomes
\be
\left(F_{\alpha_1 \alpha_2 \alpha_3 \alpha_4 \alpha_5} -  \frac12 (C_{\3} \wedge G_{\it 4})_{[\alpha_1 \alpha_2 \alpha_3 \alpha_4 \alpha_5] xy}  \right) L^{-2} \tau_2^{-1} 
\ee
which is a nine-dimensional relation. Componentwise calculation gives
\be \label{C6exp}
 C_{\3} \wedge G_{\it 4}\big|_{xy} = - B_{\it 2} \wedge F_{\it 3}+ A_{\it 2} \wedge H_{\it 3} + K_{\it 1} \wedge G_{\it 4} + A_{\it 3} \wedge H_{\it 2}.
\ee
It immediately follows that 
$$
  \left( G_{\it 7} - \frac12 C_{\3} \wedge G_{\it 4} \right)\Bigg |_{xy} = F_{\it 5}^{(10)} - \frac12 A_{\2}^{(10)} \wedge H_{\3}^{(10)} + \frac12 B_{\2}^{(10)} \wedge F_{\3}^{(10)}
 \equiv \tilde F_{\it 5}^{\rm wo(10)}$$
with all the indices different from $x,y,y'$.

Now, different components of the $F_{\it 5}$ came from different forms $\C_{\it 4}$ and $C_{\it 6}$, therefore the Lorentz covariance of the $F_{\it 5}$ is not trivial. Comparing the LHS and the RHS, we find that the covariance we need is the same coefficients
\begin{equation} \label{r}
 r =  L^{-2} \tau_2^{-1}.
\end{equation}
That is, the new degree of freedom turns out to be not independent. In terms of the vielbein, we have an emergent component
\be \label{emergenteyprime}
 e_{\underline{y'}}^{y'} = (e_{\underline{x}}^{x}e_{\underline{y}}^{y})^{-1} .
\ee
The condition (\ref{r}) can be only the necessary condition for the relation (\ref{emergenteyprime}), but later we will see that indeed we need the condition (\ref{emergenteyprime}).

Summarizing, we have {\em defined} the four form field $A_{\it 4}$ and its field strength $F_{\it 5}$ via twelve-dimensional duality relation and dimensional reduction. This is expressed as
\begin{equation}
 \tilde F ^{\rm wo (10)}_{\it 5} =  {*_{10}  \tilde F}^{\rm w(10)}_{\it 5},
 \end{equation}
where $*_{10}$ is the Hodge dual operator in ten dimension and 
\be \label{Fprime} \begin{split}
  \tilde F^{\rm w (10)}_{\it 5} &\equiv \frac{1}{4!} \tilde F_{\mu \nu \rho \sigma y'}^{(10)} dx^{\mu} \wedge dx^{\nu}  \wedge  dx^{\rho}  \wedge dx^{\sigma} \wedge  dy', \\
 \tilde F^{\rm wo (10)}_{\it 5} &= \frac{1}{5!} \tilde F_{\mu \nu \rho \sigma \tau}^{(10)} dx^{\mu}  \wedge dx^{\nu}  \wedge dx^{\rho}  \wedge dx^{\sigma} \wedge  dx^{\tau},
\end{split}
\ee
with all the indices nine-dimensional.  This only happens in this special circumstance in which the numbers of degrees are related as  
 $$ \frac12 \cdot \frac{8 \cdot 7 \cdot 6 \cdot 5}{4 \cdot 3 \cdot  2 \cdot 1} = \frac{7 \cdot 6 \cdot 5}{3 \cdot 2 \cdot 1}= \frac{7 \cdot 6 \cdot 5 \cdot 4}{4 \cdot 3 \cdot 2 \cdot 1}.
 $$
This is equivalent to ten-dimensional self-duality relation using full ten-dimensional one.

In fact, this means the unprimed version should also hold
\be \label{selfduality}
 \tilde F^{(10)}_{\it 5} = {*_{10} \tilde F}^{(10)}_{\it 5} 
\ee
where 
\be \label{fiveform10}
 \tilde F^{(10)}_{\it 5} \equiv \frac{1}{5!} \tilde F^{(10)}_{\upmu \upnu \uprho \upsigma \uptau} dx^{\upmu} \wedge dx^{\upnu}\wedge dx^{\uprho}\wedge dx^{\upsigma} \wedge dx^{\uptau} \equiv \tilde F^{\rm w (10)}_{\it 5} + \tilde F^{\rm wo (10)}_{\it 5}
\ee
 has the full dependence on ten-dimensions $\upmu, \dots, \uptau = 0,1,\dots, 9.$. The equation of motion and the Bianchi identity from this are
 \be \label{F5covariant}
\d \tilde F^{(10)}_{\it 5} = \d {*_{10} \tilde F}^{(10)}_{\it 5} = H^{(10)}_{\it 3} \wedge F^{(10)}_{\it 3},
\ee
with now the full ten-dimensional index structure.

\subsection{Couplings and scales}

Dimensional reduction of the Einstein--Hilbert term goes as follows. We compactify three dimensions using the metric (\ref{themetric})
\begin{align}
 \sqrt{-\G} {\cal R}  
  & = \sqrt{-g'_{(9)}} r L^2 \tau_2 \left( R_{(9)} - 2 (L \tau_2)^{-1} \nabla^2 (L\tau_2) - \frac12 \tau_2^{-2} (\partial_\mu \tau_1)^2 - 2L^{-1} \nabla^2 L  \right. \nn \\
  & \quad   \left.  - \frac14 L^2 \tilde f_{\mu \nu}^2 - \frac14 L^2 \tau_2^2 h_{\mu \nu}^2  \right).
\end{align}
Here, $\tilde f_{\mu \nu} \equiv f_{\mu \nu} - A h_{\mu \nu}$.
The overall factor becomes $r L^2 \tau_2 =1$ by (\ref{r}). The last two terms provides the kinetic terms for the dimensionally reduced components of $\tilde F_{\mu \nu y'}$ and $H'_{\mu \nu y'}$ in (\ref{oxidation}).
Using the identity (\ref{identity}) in the appendix, this can be recollected as
\be
 \sqrt{-g'_{(9)}} \left( R_{(9)} - 2 r^{-1} \nabla r + 4 \tau_2^{-2} (\partial_\mu \tau_2)^2  - \frac14 \tau_2^{-2} (\partial_\mu \tau_1)^2 \right).
\ee
With the term (\ref{Kkinetic}), this is nothing but the dimensional reduction of ten-dimensional type IIB action
\be
 \sqrt{-g_{(10)}} r^{-1} \left( R_{(10)} + \frac{4}{\tau_2^{2}} (\partial_\upmu \tau_2)^2 - \frac{1}{4 \tau_2^2} (\partial_\upmu \tau_1)^2 \right).
\ee
with the decompactified metric (\ref{recoveredmetric}).

Identifying that $\tau_2= g_{\rm IIB}^{-1}$ we would require $L$ should be absent in the ten-dimensional IIB action, which could be expected by counting degrees of freedom. We can show that the only way to remove $L$ in the action is to  absorb $r = L^{-2} \tau_2^{-1}$ in the coupling and to rescale
\be \label{rescaling}
 g_{\upmu \upnu}^{\prime (10)} \equiv L^{-1} g_{\upmu \upnu}^{ (10)}.
\ee
Accordingly, we have $\sqrt{-g'_{(10)}} = \sqrt{-g_{(10)}} L^{-5},$ and $R^{(10)} \to L R^{(10)}$ up to a kinetic term for $L$. This fixes the ten-dimensional IIB gravitational coupling in terms of the eleven-dimensional coupling $\kappa_{11}$, which may be a more fundamental quantity than $\kappa_{12}$,
\begin{equation} \label{IIBcoupling}
\frac{1}{2 \kappa_{\rm IIB}^2} =  \frac{(2 \pi \ell  )^2  \langle r \rangle  }{2 \kappa_{12}^2} = \frac{2 \pi \ell    }{2 \kappa_{11}^2}.
\end{equation}
This is also useful in the decompactification limit $r \to \infty$, since this can be taken as $L \to 0$ while keeping $\kappa_{11}$ fixed and the IIB coupling should be free parameter.

The rescaling (\ref{rescaling}) should also rescale the coordinate periodicity as
\be \label{stringlength}
  \ell \to  L^{-1/2} \ell \equiv \ell_{\rm s},
\ee
in which unit we can naturally convert between IIA and IIB theories in ten dimensions. The relation between the two radii from (\ref{recoveredmetric}) in the new unit are now 
\be
 R_{y}  = L^{3/2} \tau_2 \ls, \quad R_{y'} = \frac{ \ls}{L^{3/2} \tau_2} = \frac{\ls^2}{R_{y}}.
\ee

Finally we reduce the Chern--Simons term (\ref{CSterm}). Since this coupling is topological we do not care about the metric and we keep the general indices. First we note that $\G_{\it 5}$ should always have an index on $y'$, becoming $F_{\it 5}^{\rm w}$.  The other part, $C_{\3} \wedge G_{\it 4}$, should have two induces on $x$ and $y$. It has exactly the expansion (\ref{C6exp}). Noting that $\int_{T^2}  dx \wedge dy= (2 \pi \ell)^2$, we obtain
\begin{align} 
 S_{\rm CS} =
 & \frac{ (2 \pi \ell)^2}{4 \kappa_{12}^2} \int_{10} r F_{\it 5}^{(9)\rm w} 
 \wedge ( B_{\it 2}^{(9)} \wedge F_{\it 3}^{(9)}- A_{\it 2}^{(9)} \wedge H_{\it 3}^{(9)} - K_{\it 1}^{(9)} \wedge G_{\it 4}^{(9)} - A_{\it 3}^{(9)} \wedge H_{\it 2}^{(9)}) \nn \\ 
 = & \frac{1 }{4\kappa_{\rm IIB}^2} \int F_{\it 5}^{(10) \rm w} \wedge ( B^{(10)}_{\it 2} \wedge F^{(10)}_{\it 3} - A^{(10)}_{\it 2} \wedge H^{(10)}_{\it 3}) \\
 =& \frac{1}{2\kappa_{\rm IIB}^2} \int F_{\it 5}^{(10)\rm w} \wedge  B^{(10)}_{\it 2} \wedge F^{(10)}_{\it 3}. \nn
\end{align}
The integration is done over the remaining ten dimension including the $y'$-direciton. We performed integration by parts in the last line.

\subsection{Ten-dimensional covariant action}

We are now in a position to write down the action of type IIB supergravity in ten dimension. We obtained it using compactification and also decompaction of already compact direction. If we give up covariant formulation, nevertheless the Poincar\'e symmetry is to be recovered, the ten-dimensional action can be written as dimensionally reduced form from the twelve-dimensional one as follows.
\begin{align}
-\frac{1}{4\kappa_{12}^2} & \int d^{12} x \sqrt{-\G}  \,\G_{\alpha \beta \gamma\delta y'}^2  =  - \frac{1}{4 \kappa_{\rm IIB}^2}  \int d^{10} x \sqrt{-g}  |\tilde F_{\it 5}^{\rm w}|^2  , \\
-\frac{1}{4\kappa_{12}^2} &\int d^{12} x \sqrt{-\G}  \G_{\alpha \beta \gamma x y'} ^2 = -\frac{1}{4\kappa_{\rm IIB}^2} \int d^{10} x \sqrt{-g} \tau_2^2 \left(|H_{\3}|^2 - \frac12 h_{\alpha \beta }^2 \right), \\
-\frac{1}{4\kappa_{12}^2} &\int d^{12} x \sqrt{-\G}  \G_{\alpha \beta \gamma y y'}^2  = -\frac{1}{4 \kappa_{\rm IIB}^2} \int d^{10}x  \sqrt{-g} \left(|\tilde F_{\3}|^2- \frac12 \tilde f_{\alpha \beta }^2 \right), \\
 \frac{1}{12\kappa_{12}^2} & \int \C_{\it 4} \wedge G_{\it 4} \wedge G_{\it 4}= \frac{1}{2 \kappa_{\rm IIB}^2} \int F_{\it 5}^{\rm w} \wedge B_{\2} \wedge F_{\3}, \label{FBFterm} \\
 \frac{1}{2\kappa_{12}^2} & \int d^{12} x \sqrt{-\G} \left({\cal R}-\frac12 \G_{\alpha \beta x y y'}^2 \right) \label{EHreduction} \\
& = \frac{1}{2\kappa_{\rm IIB}^2} \int d^{10} x \sqrt{-g}  \tau_2^2\left( R +4 (\partial_\mu \tau_2)^2- \frac12 |F_{\1}|^2 
   - \frac14 h_{\alpha \beta}^2 - \frac14 \tilde f_{\alpha \beta}^2 \right). \nn
\end{align}
All the fields here are ten-dimensional, so we suppress the superscript $^{(10)}$ appearing in those fields in the appendix. In particular $\tilde F^{\rm w}_{\it 5}$ is defined in (\ref{Fprime}). Also, $g$ is the determinant of the ten-dimensional metric, with which we calculate the Ricci scalar $R$. We have defined 
\begin{align}
 \tilde F_{\it 3} &= F_{\3} - A \wedge H_{\3}, \quad F_1 = dA .
\end{align}

It is remarkable that, we do not lose ten-dimensional Poincar\'e invariance only with $\tilde F_5^{\rm w}$, as long as it satisfies self-duality condition (\ref{selfduality})
\be
 |\tilde F_{\it 5}^{\rm w}|^2 = \frac12 |\tilde F_{\it 5}^{\rm w}|^2 + \frac12 |{\tilde F}^{\rm wo}_{\it 5}|^2 = \frac12 |\tilde F_{\it 5}|^2.  
\ee
This is possible because either field contains the full physical degrees of freedom and satisfies the relation $|\tilde F_{\it 5}^{\rm w}|^2 = |\tilde F_{\it 5}^{\rm wo}|^2$ from the Hodge duality. 
The last term has manifest covariance, but after plugging in the action, the normalization has an extra $\frac12$ factor, as a footprint of self-dual degree of freedom.

The Chern--Simons term (\ref{FBFterm}) has a special property
$$
 \int F^{\rm w}_{\it 5} \wedge B_{\2} \wedge F_{\3} = \int \left( F^{\rm w}_{\it 5} -  \frac12 B_{\2} \wedge F_{\it 3} + \frac12 A_{\it 2} \wedge H_{\it 3} \right) \wedge B_{\2} \wedge F_{\3} = \int \tilde F^{\rm w}_{\it 5} \wedge B_{\2} \wedge F_{\3}
$$
where we have used the equality $F_{\3} \wedge F_{\3} = 0$ and done integration by parts. Use the definition (\ref{F5covariant}) $ \tilde F^{\rm w}_{\it 5} + \tilde F^{\rm wo}_{\it 5} = \tilde  F_{\it 5}$ and totally antisymmetric property $\tilde F^{\rm wo} \wedge B_{\2} \wedge F_{\3} = 0.$
We have
\be
 \frac{1}{12\kappa_{12}^2} \int \C_{\it 4} \wedge G_{\it 4} \wedge G_{\it 4}= \frac{1}{4 \kappa_{\rm IIB}^4} \int F_{\it 5} \wedge B_{\2} \wedge F_{\3}, 
\ee
with the full ten-dimensional invariance.
We have finally recovered ten-dimensional Poincar\'e invariance in the action, but the self-duality condition is lost. We have to impose the condition after having equation of motion.

\section{The fermionic part and supersymmetry} \label{sec:fermonic}

In this section we embed the fermonic sector and look for supersymmetry transformations. It turns out that the fermionic degrees of freedom of eleven-dimensional supergravity suffice, because the components along the twelfth direction becomes emergent, as in the case of the graviton. 

The most beneficial feature of twelve dimensional, and in general of $4n$-dimensional, embedding is, we can exchange among Majorana and Weyl spinors of all the chirailities just using the property of the spacetime. Therefore we can exchange between parity preserving IIA supersymmetry generators with chiral IIB ones in twelve-dimensions.

\subsection{Embedding in twelve dimension}

We first summarize the properties of spinors for agreement of notations. The ten-dimensional Clifford algebra is generated by ten matrices $\upgamma^\upalpha$ satisfying 
\be \label{Clifford}
 \{ \upgamma^\upalpha,\upgamma^\upbeta \} = 2 \eta^{\upalpha \upbeta} {\bf 1}_{32}, \quad \eta^{\upalpha \upbeta} = {\rm diag\,}(-1,1,1,\dots, 1)
\ee 
with the rank-32 unit matrix ${\bf 1}_{32}$. Under this a Dirac spinor $\psi$ transforms and can be expressed in terms of two 16-complex-component spinors  $\alpha$ and $\beta$ as
\be \label{10Dspinor}
 \begin{pmatrix} \psi_1 \\ \psi_2 \end{pmatrix},
\ee
to make up 32 components. Each component is nine-dimensional Dirac.

We may impose Weyl or Majorana conditions, each of which reduces half the components. In ten dimension, both conditions can be imposed at the same time.
The Weyl spinors are eigenstates of the chirality operator 
\be \label{gamma10}
 \upgamma^{10} = \prod_{\upalpha=0}^{9} \upgamma^\upalpha =  \begin{pmatrix}  {\bf 1} & 0  \\ 0 & {\bf -1}    \end{pmatrix} ,
\ee
where ${\bf 1}$ is the rank-16 unit matrix in a certain basis. We always have such a basis. The Majorana spinor satisfies the `reality' condition
\be \label{Majcond}
 \begin{pmatrix}  \psi_1^* \\  \psi_2^*  \end{pmatrix} =  \begin{pmatrix}  \psi_1 \\  \psi_2  \end{pmatrix}.
\ee

The spinor (\ref{10Dspinor}) can be readily promoted to an eleven-dimensional Dirac spinor having now the dependence on $x^m$, if we use $\gamma^{10}$ in (\ref{gamma10}) as the eleventh generator of the algebra  (\ref{Clifford}). In this case, only the Majorana condition (\ref{Majcond}) can be imposed.  

Since we have an extra coordinate $y'$, we should go to twelve-dimension and need another gamma matrix extending the Clifford algebra. Therefore we attempt to embed it in twelve-dimension, in which  a spinor has 64 components.
We do this by extending the gamma matrices as
\be \label{gammaext}
 \Gamma^\upalpha = \begin{pmatrix} - \upgamma^\upalpha &   \\  &  \upgamma^\upalpha \end{pmatrix}, \upalpha=0,\dots,9, \quad 
 \Gamma^{x} = \begin{pmatrix} & & {\bf 1} &   \\ & &  & {\bf 1}  \\ {\bf 1} &   & & \\  & {\bf 1} & & \\ \end{pmatrix}, \quad
 \Gamma^{y'} = i \begin{pmatrix} & & -{\bf 1} &   \\ & &  & -{\bf 1}  \\ {\bf 1} &   & & \\  & {\bf 1} & & \\ \end{pmatrix}, \quad
\ee
where $\upgamma^\upalpha$ are ten-dimensional gamma matrices. One can see that $\Gamma^\upalpha, \Gamma^x, \Gamma^{y^{\prime}}$ satisfy the commutation relation in a similar manner as (\ref{Clifford}). We cannot simply embed {\em eleven}-dimensional spinor $\psi$ in twelve dimension, because we did not extended $\upgamma^{10}$ to the off-diagonal block matrix $\Gamma^{x}$, unlike other $\upgamma^\alpha$'s defined in the first of (\ref{gammaext})
$$
 \begin{pmatrix}  - \upgamma^{10} &   \\  &  \upgamma^{10} \end{pmatrix}  \ne \Gamma^{x} .
$$
Interestingly but not coincidentally the left-hand side is the twelve-dimensional chirality operator
\be \label{embeddingcond}
 \begin{pmatrix}  - \upgamma^{10} &   \\  & \upgamma^{10} \end{pmatrix} = \Gamma^{x} \Gamma^{y'} \prod_{\upalpha=0}^{9} \Gamma^\upalpha =-  \begin{pmatrix}  {\bf 1} & & &  \\ &- {\bf 1} & &  \\ & & -{\bf 1} &  \\  & & & {\bf 1}  \\ \end{pmatrix} \equiv -\Gamma.
\ee
The remedy follows, if we require $\Gamma^x \Psi$ be compatible to ten-dimensional $\Gamma \Psi$ by making them proportional. Simple identification, however, does not give us a nontrivial embedding. A good choice is
\be
 \Gamma^x \Psi =i \Gamma \Psi.
\ee
This reduces the number of components to be half
\be \label{11spinorin12}
\Psi = \begin{pmatrix}  \psi_1 \\ \psi_2 \\ i \psi_1 \\ -i  \psi_2 \end{pmatrix}.
\ee
This shows how do we embed the eleven-dimensional spinor (\ref{10Dspinor}) into twelve dimension. Conversely, the twelve-dimensional spinor (\ref{11spinorin12}) decomposes into two spinors (\ref{10Dspinor}) and $i \upgamma^{10}$ times it, each of which describes the same dynamics, governed by the same equation.

We emphasize again that this spinor is an eleven-dimensional one embedded in twelve-dimensions, with the eleven-dimensional Poincar\'e symmetry and Lorentz signature $(1,11)$. So we can reduce half of the components by the property of the eleven dimension. In fact, imposing twelve-dimensional Majorana condition (\ref{12DMajorana}) on the the spinor (\ref{11spinorin12}) reduces it to be eleven-dimensional Majorana, which is again interpreted as ten-dimensional Majorana, with purely real $\psi_1$ and $\psi_2$.

Finally, it is always possible to name a coordinate to be $y$ such that the corresponding $\gamma$ matrix is
\be  \label{gammay}
 \upgamma^y =i \begin{pmatrix} 0 & -{\bf 1} \\ {\bf 1} & 0 \end{pmatrix}, \quad \Gamma^y = \begin{pmatrix} -\upgamma^y & 0 \\  0 & \upgamma^y \end{pmatrix}=
 i \begin{pmatrix} & {\bf 1} & & &  \\ -{\bf 1} & & &  \\ & & &-{\bf 1} \\ &  & {\bf 1}  & \\ \end{pmatrix},
\ee
where the latter has a doubled dimension.
In the bosonic part, we have compactified this $y'$ direction to have nine-dimensional action. This does a special operation
\be
\upgamma^y \begin{pmatrix}  \psi_1 \\ \psi_2\end{pmatrix} = \begin{pmatrix} -i  \psi_2 \\ i \psi_1 \end{pmatrix}
\ee
exchanging the chirality of ten-dimensional Weyls. It is possible after compactifying the $y$-direction, because we do not have ten-dimensional chirality in nine-dimensions.

\begin{figure}[t]
$$
\xy
\xymatrixcolsep{6pc}
\xymatrixrowsep{2.4pc}
\xymatrix{
  *+{\text{12D}\begin{pmatrix} \psi_1 \\ \psi_2 \\ i \psi_1 \\ - i \psi_2 \end{pmatrix} \Leftrightarrow \begin{pmatrix} 0 \\ \psi_2 \\ i \psi_1 \\ 0 \end{pmatrix}}  \ar[r]_{\blue \quad S^1_{y'} \text{compactification}} \ar[d]^{T^2 \simeq S^1_x \times S^1_{y}\text{ comp.}} 
   &*+{\text{11D} \begin{pmatrix} \psi_1 \\ \psi_2 \end{pmatrix} } \ar[d]^{S^1_{x} \text{ comp.}}  \\
*+{\text{IIB}   \begin{pmatrix} 0 \\ i \psi_1 \end{pmatrix} + \begin{pmatrix} 0 \\ \psi_2 \end{pmatrix} }  \ar@{<-->}[r]_{\text{\blue $T$-dual}} & *{\text{ IIA}  \begin{pmatrix} \psi_1\\ 0 \end{pmatrix} 
+ \begin{pmatrix} 0 \\  \psi_2\end{pmatrix}  }
}
\endxy
$$
\caption{Reduction of an eleven-dimensional Majorana spinor {\em embedded in twelve dimension}. In twelve dimension, a Majorana spinor is converted to Weyl spinor. Going to nine dimension and performing the decompactification, the gamma matrix structure forces the resulting total spinor to be a complex Weyl in ten dimension. Each entry $\psi_1$ or $\psi_2$ is nine-dimensional Majorana spinor having 16 real components.} \label{f:spinorreduction}
\end{figure}

\subsection{Generalized local supersymmetry transformations}

We start with the Rarita--Schwinger action for the gravitino $\Psi_m$
$$
 S_{\rm RS} = - \frac{1}{4\kappa_{12}^2 }\int d^{11}x dy'  \sum_{m,n,p=0}^{10} i \overline{\Psi}_m \Gamma^{mnp} \partial_n \Psi_p ,
$$
with the usual Dirac conjugate $\overline \Psi_m =\Psi_m^\dagger \Gamma^0$ with an appropriate choice of $\Gamma^0$ and and antisymmetrized $\Gamma$-matrices
$$ 
\Gamma^{mnp} = \Gamma^{[m} \Gamma^{n} \Gamma^{p]}.
$$
We have only eleven-dimensional Poincar\'e invariance in the beginning. Since we have only 11 derivative terms, time evolution does not ruin the embedded form, and the physics is the same as eleven-dimensional supergravity.
Therefore, from a twelve-dimensional spinor, which is not fully Lorentz covariant but is only under eleven-dimensional, we can obtain the following by truncation to ten dimension. 

We require the {\em eleven-}dimensional local supersymmetry, expressed in terms of the spinors and gamma matrices embedded in the twelve dimension as above. Then keeping the half of the components, 
\be \label{SUSYMajorana}
 {\cal E} = \begin{pmatrix} \epsilon_1 \\ \epsilon_2 \\ i \epsilon_1 \\ -i \epsilon_2 \end{pmatrix} \to \begin{pmatrix} \epsilon_1 \\ \epsilon_2 \end{pmatrix}, \quad \epsilon_1,\epsilon_2 \text{ real},
\ee
we obtain the eleven-dimensional supergravity.
Therefore the natural extension is
\begin{align}
 \delta e^{a}_m &= \frac12 \bar {\cal E} \Gamma^a \Psi_m \\
  & = \bar \epsilon \Gamma^a \Psi_m \nn \\
 \delta \C_{mnp \underline{y'}} &= -\frac{3 r}{2} \bar {\cal E} \Gamma_{[mn} \Psi_{p]} = - \frac12 \bar {\cal E} \Gamma^{y'} \Gamma_{[\underline{y'}mn} \Psi_{p]} \\
  & = - 3 \bar \epsilon \Gamma_{[mn} \Psi_{p]}, \nn \\
 \delta \Psi_m &= \left( \nabla_m  + \frac{1}{288} ({\G}_{npqr\underline{y'}}\Gamma_m \Gamma^{npqr}  -12 {\G}_{mnpq\underline{y'}} \Gamma^{npq}  )\right) {\cal E} \label{SUSYgravitino} \\
  & \to \left( \nabla_m + \frac{1}{288} ( G_{npqr}\Gamma_m  \Gamma^{npqr} - 12 G_{mnpq} \Gamma^{npq} )\right) \epsilon. \nn
\end{align}
The only difference is coordinate dependence, that is all $f(x^m)$ are changed as $f(x^m,y)$, while the tensor structure is intact. In this section, the underlined coordinate values are that of general spacetime, whereas plane coordinates are local flat ones. All the indices are eleven-dimensional, different from $y'$, and $\bar {\cal E} ={\cal E}^\dagger \Gamma^0$.
We note that although the degree is enhanced, there is no extra transformation than those of eleven-dimensional supergravity.
The transformation of zw\"olfbein takes the form
\be \label{SUSYzwoelfbein}
 \delta e^{\alpha}_\mu = \frac12 \bar {\cal E} \Gamma^\alpha \Psi_\mu, \quad
 \delta e^x_{\underline{x}}  = \frac12 \bar {\cal E} \Gamma^{x} \Psi_{\underline{x}}, \quad
 \delta e^y_{\underline{y}}  = \frac12 \bar {\cal E} \Gamma^{y} \Psi_{\underline{y}}, 
\ee
which are merely re-expressing the original rules. 

We have seen in (\ref{emergenteyprime}) that the components of the graviton in the $y'$-direction emerge from a combination of components in the other directions. Likewise, we do not require the presence of the component $\Psi_{y'}$ along the twelfth direction, but it will emerge later after decompactification. In fact, a would-be independent component $\Psi_{y'}$ would have resulted in an extra component of graviton whose helicity exceeds 2 in the four dimensional language, which is inconsistent.

\subsection{Supersymmetry transformation in nine dimension}

Now we go to nine dimension by compactfying three dimensions as before with the metric (\ref{themetric}). Eventually we decompactify some of dimensions to obtain ten-dimensional supergravity.  As in the bosonic case, we have two different but equivalent presentation of supergravities in nine dimension, which decompactify in ten dimension to IIA or IIB supergravity. 

We consider ten-dimensional Poincar\'e symmetry including the $y'$-direction to have IIB supergravity. We now choose a supersymmetry generating spinor as Weyl
\be \label{12DWeyl}
 {\cal E}=  \begin{pmatrix} \epsilon_1 \\ 0 \\ 0 \\ - i \epsilon_2  \end{pmatrix}, 
\ee
where each $\epsilon_1,\epsilon_2$ has 16 real components.
Since each $\Gamma^x$, $\Gamma^y$, or $\Gamma^{y'}$ anticommutes the rest of the gamma matrices, we can push it to the rightmost to directly act on $\cal E$. 
In this explicit form, we can act any gamma matrices. Since we have nine-dimension, in fact the components form linear combinations. For example,
\be
 {\cal E} \to \epsilon_1- i \epsilon_2 \equiv \ve^*, \quad \Gamma^{y'} {\cal E} = \begin{pmatrix} 0 \\ -\epsilon_2 \\  i \epsilon_1 \\ 0 \end{pmatrix} \to 
  i \epsilon_1 - \epsilon_2 = i \ve , 
\ee
where we have combined two real Majorana spinors into one complex in nine dimension. Similar relations are shown in (\ref{Gammaaction}) in the appendix. This shows how the twelve-dimensional structure selects the spinor, which  cannot be seen in ten dimensions. 

We have essentially the same form for the transformation of the antisymmetric tensor fields in ten dimensions. 
Wae investigate the transformations of the gravitinos
\begin{align}
  L^{-1/2} \delta \psi_\alpha  & = \left(E^{\mu}_{\alpha} \nabla_\mu - \frac{L^{-1}}{4} {\gamma_\alpha}^\mu \partial_\mu L 
  - \frac{iL^{3/2}}{24}  (3 \tilde {\mathbb F}_{\it 5}^{\rm w} \gamma_\alpha + \gamma_\alpha \tilde {\mathbb F}_{\it 5}^{\rm w})\right)  \ve  \nn \\
 & 
 \quad +\left(- \frac{iL^{3/2}}{8}  ( \tilde {\mathbb F}_{\it 3}^{\rm w} \gamma_\alpha - \gamma_\alpha \tilde {\mathbb F}_{\it 3}^{\rm w})
  + \frac{L^{3/2} \tau_2}{8} ({\bf h}_{\2} \gamma_\alpha - \gamma_\alpha {\bf h}_{\2})
  \right. \label{Psialpha} \\
 &
  \left. \quad + \frac{1}{24} (3 \tilde {\mathbb H}_{\it 3} \gamma_\alpha - \gamma_\alpha \tilde {\mathbb H}_{\it 3}) + \frac{1}{24 L^{3/2}\tau_2 } (3{ \mathbb H}_{\2} \gamma_{\alpha}  - \gamma_\alpha{\mathbb H}_{\2} )  
  - \frac{i}{24 \tau_2}  (3 \tilde {\mathbb F}_{\it 3}  \gamma_\alpha + \gamma_\alpha \tilde {\mathbb F}_{\it 3}) \right)   \ve^* . \nn \\
 L^{-1/2}\delta \psi_x   &
 = \left(-  \frac{L^{-1}}{2} \gamma^\mu \partial_\mu L   -\frac{iL^{3/2}}{12}  \tilde {\mathbb F}_{\it 5}^{\rm w} \right) \ve 
 +\left(  \frac{1}{6}  \tilde{\mathbb H}_{\it 3} 
    +  \frac{i}{6 L^{3/2}\tau_2 } { \mathbb H}_{\2}
  - \frac{i}{12 \tau_2}  \tilde {\mathbb F}_{\it 3} \right) \ve^*. \\
 L^{-1/2}  \delta \psi_y & = -\left(  \frac{  L^{-1}}{2} \gamma^\mu \partial_\mu L
 + \frac{1}{4}  {\mathbb F}_{\1}
 +\frac{iL^{3/2}}{12}  \tilde {\mathbb F}_{\it 5}^{\rm w} \right) i \ve 
  +\left(
  - \frac{1}{12} \tilde {\mathbb H}_{\it 3} + \frac{1}{6 L^{3/2}\tau_2 } { \mathbb H}_{\2}   
 + \frac{i}{6\tau_2}  \tilde {\mathbb F}_{\it 3} \right) i\ve^*. 
 \end{align}
Here the decomposition becomes straightforward in the local flat coordinates.
Also ${\gamma_{\alpha}}^\mu = \frac12[\gamma_\alpha, \gamma^\mu]$. In this section, Greek indices are nine-dimensional. 

Using the result of Subsection \ref{s:reduction}, summarized in (\ref{G5expA}-\ref{ExptoF3}) in the appendix, we reduce the five-form field strength in (\ref{SUSYgravitino}) as
 \begin{align}
 \frac{1}{4!} \G_{abcd y'} \Gamma^{abcd}   = L^2 \tilde {\mathbb F}_{\it 5}^{\rm w}  \Gamma^{y'} +  L^{1/2} \tilde {\mathbb H}_{\3} \Gamma^x + L^{1/2} \tau_2^{-1} \tilde {\mathbb F}_{\3} \Gamma^y + L^{-1} \tau_2^{-1} {\mathbb H}_{\2}\Gamma^x \Gamma^y 
\end{align}
where we used $(\Gamma^A)^2=1$ for $A=x,y,y'$. Note that we have a chirality flipping operator $\Gamma^{y}$ from the ten-dimensional point of view. We have straightforward definitions
\begin{align}
\tilde {\mathbb F}_{\it 5}^{\rm w} &\equiv  \frac{1}{5!} \tilde F_{\alpha \beta \gamma \delta y'} \gamma^{\alpha \beta \gamma \delta y'} =  \frac{1}{4!} \tilde F_{\alpha \beta \gamma \delta y'} \gamma^{\alpha \beta \gamma \delta }\gamma^{y'}    , \\
 \tilde {\mathbb F}_{\3} & \equiv \frac{1}{3!} \tilde F_{\alpha \beta \gamma } \gamma^{\alpha \beta \gamma }, 
\end{align}
and similar definitions are given in (\ref{F1})-(\ref{F3w}) in the appendix.
Going down to nine-dimension, we need to keep only the $16\times 16$ block of the gamma matrices, thus we then replace all $\Gamma^\alpha$ to $\gamma^\alpha$ and their antisymmetrized. Although each gamma matrix flips the chirality, we always have even number of gamma matrices, so that we have only one chirality thus far.

\subsection{Emergent component of the gravitino}

Although we did not have the  $y'$-component of the gravitino in the original theory, this may arise as follows. First, we have redundant dimensional relation for $e_{\underline{y'}}^{y'}$ in (\ref{emergenteyprime}). From its supersymmetry variation, we define an auxiliary gravitino component $\Psi_{y'}$
\be  \label{Psiyprimedef}
 \delta e^{y'}_{\underline{y'}}  = - \frac{1}{L^3 \tau_2^2} (\tau_2 \delta e_{\underline{x}}^x + \delta e_{\underline{y}}^y)  \equiv \frac12  \bar{\cal E} \Gamma^{y'} \Psi_{\underline{y'}}.  
\ee 
By the transformations of the other fields in (\ref{Psiyprimedef}) that we already know in (\ref{SUSYzwoelfbein}), we have effectively defined a new component
\be
  \Psi_{\underline{y'}}  = - \frac{1}{L^3 \tau_2^2} (\Gamma^{y'})^{-1} (\tau_2 \Gamma^x \Psi_{\underline{x}}  + \Gamma^y \Psi_{\underline{y}}).
\ee
Going to local Lorentz coordinates, we get a simpler relation
\be
  \Psi_{y'} =  - \Gamma^{y'} (\Gamma^x \Psi_x + \Gamma^y \Psi_y),
 \ee
using $(\Gamma^{y'})^{-1} = \Gamma^{y'}.$ In fact, we do not have covariance along each direction, but this does not matter because all the directions involved here are compact. 
Rewriting the form making the $y'$ dependence explicit, we have exactly the same transformation as the other coordinate, that is Eq. (\ref{Psialpha}) with the index $\alpha$ replaced by $y'$. Note that because of the even number of gamma matrices, we have the same chirality. 

Here is where the supersymmetry comes into play. In the bosonic sector, ten-dimensional Lorentz invariance in IIB space including the $y'$-direction forced the degree of freedom $e_{y'}^{y'}$ to not be independent as in (\ref{oxidation}). The supersymmetry related this field to $\Psi_{y'}$. Therefore, we have the same bosonic degrees of freedom in twelve dimension as those of eleven-dimensional supergravity. All the fields in the twelfth direction are emergent, along with the graviton in (\ref{emergenteyprime}). This circumvents the no-go theorem by Nahm \cite{Nahm:1977tg}.

\subsection{The IIB supergravity transformations} 

We go to ten dimension by decompactifying the $y'$-direction, whose configuration is shown in Figure \ref{f:coordinates}. To obtain the conventional form, we modify the definition on gravitnos and dilatinos and perform dimensional reduction again 
\be
\psi_\alpha^{\rm new} \leftarrow L^{-1/4} \left(  \Psi_\alpha + \frac12 \Gamma_\alpha \Gamma^y \Psi_y  \right), \quad 
\lambda^{\rm new} \leftarrow L^{-1/4} \left(\Gamma^x \Psi_x - \Gamma^y \Psi_y \right), \quad
\ve^{\rm new} \leftarrow L^{-1/4} \ve
\ee
The application of $\Gamma^y$ puts an extra $i$ in $\Psi_y$.
We will drop the superscript $^{\rm new}$ in what follows without confusion.
The dilatino $\lambda$ is the superpartner of the dilaion $\tau_2$ and this redefinition extracts the component.

Going to general coordinates, we have
\begin{align}
 \delta \lambda  & = - \frac12 \gamma^\mu \left(L^{-1} \partial_\mu L + i \partial_\mu A \right) 
   \ve   \\
 & + \frac14 \left(\frac{i}{\tau_2} \tilde {\mathbb F}_{\it 3}^{\rm wo}  + 
i L^{3/2}{\mathbb F}_{\3}^{\rm w}   -  {\mathbb H}_{\it 3} - L^{3/2}\tau_2{\bf h}_{\it 2} \right)  \ve^*,\nn \\ 
  \delta \psi_\mu   &=  \left(\nabla_\mu - \frac{L^{-1}}{4} \gamma_\mu \gamma^\nu \partial_\nu L + \frac{i}{8} \gamma_\mu {\mathbb F}_{\1}  + \frac{i L^{3/2}}{8 } \tilde {\mathbb F}_{\it 5}^{\rm w} \gamma_\mu \right)  \ve \label{Psimu9d} \\
&   + \frac{1}{8} \left( \gamma_\mu {\mathbb H}_{\3} - {\mathbb H}_{\3} \gamma_\mu - {L^{3/2} \tau_2} ({\bf h}_{\2} \gamma_\mu - \gamma_\mu {\bf h}_{\2} ) 
 - \frac{i}{\tau_2} \tilde {\mathbb F}_{\it 3}^{\rm wo} \gamma_\mu 
 - i L^{3/2}{\mathbb F}_{\3}^{\rm w}\gamma_\mu \right)   \ve^*, \nn
 \end{align}
and essentially the same term in the $y'$ direction as the last (\ref{Psimu9d}). 
Using the results of the previous section, we obtain ten-dimensional fields out of the nine-dimensional ones
 \begin{align}
 L^{3/2}  \tilde {\mathbb F}^{\rm w}_{\it 5}  &=   \frac{1}{ \tau_2} \tilde {\bf F}_{\it 5}^{\rm w (10)} \equiv \frac{1}{5!}\tilde F^{(10)}_{\mu \nu \rho \sigma y'} \upgamma^{\mu \nu \rho \sigma y'}, \\ 
  \tilde {\mathbb H}_{\it 3} + L^{3/2}\tau_2{\bf h}_{\it 2} \gamma^{y'} &= \tilde {\bf H}_{\3}^{(10)} \equiv \frac{1}{3!} \tilde H^{(10)}_{\mu \nu \rho} \upgamma^{\mu \nu \rho}, \label{H3decompact} \\
  \frac{1}{\tau_2}\tilde{\mathbb F}_{\it 3} + L^{3/2} {\mathbb F}_{\it 3}^{\rm w} &= \frac{1}{\tau_2} \tilde {\bf F}_{\3}^{(10)}\equiv \frac{1}{3!} \tilde F^{(10)}_{\mu \nu \rho} \upgamma^{\mu \nu \rho}.
 \end{align}
noting that $r = L^{-3/2} \tau_2^{-1}$. Note again in (\ref{H3decompact}), ${\bf h}_{\2} = \partial_\mu b_\nu \Gamma^{\mu \nu}$ came from KK graviton from eleven-dimensional supergravity point of view, not from $\C_4$. But it becomes the component of ten-dimensional Kalb--Ramond field of IIB supergravity in the decompactification limit, as seen in (\ref{oxidation}). Also note that $\tilde {\mathbb H}_{\3}$ can be rewritten as
$$
 \tilde {\mathbb H}_{\3} \equiv \frac{1}{3!} (H_{\alpha \beta \gamma } + 3 K_{[\alpha}  h_{\beta \gamma]}) \gamma^{\alpha \beta \gamma },
$$
using the relation (\ref{recovery}). Also, we have decompactification of gravitons
$$
 E^\upmu_\upalpha \nabla_\upmu \ve=  \left( E^\mu_\alpha \nabla_\mu  +  \frac{1}{4 L^{3/2}\tau_2 } ({\mathbb H}_{\2}^{\rm wo} \gamma_{\alpha}  - \gamma_\alpha{\mathbb H}_{\2}^{\rm wo} )  \right)   \ve.
$$

Now we consider self-duality of four form field.
We have Hodge-like duality for gamma matrices in twelve dimensions, 
\be
 \Gamma^{M_1 \dots M_5} \Gamma =  \frac{1}{7!}  \epsilon^{M_1 \dots M_{12}} \Gamma_{M_6 \dots M_{12}}.
\ee
Fixing one index on the left-hand side to be $y'$ and two of the indices to be $x$ and $y$ on RHS also gives ten-dimensional relation
\be
 \upgamma^{\mu_1 \dots \mu_4 \underline{ y'}} \upgamma^{10} = \frac{1}{5!} \epsilon^{\mu_1 \dots \mu_{10}} \upgamma_{\mu_6 \dots \upmu_{10}},
\ee
where $\upgamma$-matrices are ten-dimensional but the indices do not take $y'$. This means,
\be \begin{split}
 {\bf \tilde F}_{\it 5}^{\rm w} &= \frac{1}{5!} \tilde F_{\mu_1 \dots \mu_4 \underline{ y'}} \upgamma^{\mu_1 \dots \mu_4 \underline{ y'}} \upgamma^{10} \\
  &= \frac{1}{(5!)^2} \tilde F_{\mu_1 \dots \mu_4 \underline{y'}}\epsilon^{\mu_1 \dots \mu_4 \underline{ y'}\mu_6 \dots \mu_{10}} \upgamma_{\upmu_6 \dots \upmu_{10}} \\
   &=  \frac{1}{5!} \tilde F^{\mu_6 \dots \mu_{10}} \upgamma_{\mu_6 \dots \mu_{10}} \\
   &={\bf \tilde F}_{\it 5}^{\rm wo}.
\end{split}
\ee
Therefore we again recover covariant form
\be
 \tilde {\bf F}_{\it 5}^{\rm w} = \frac12 \tilde {\bf F}_{\it 5}^{\rm w} + \frac12 \tilde {\bf F}_{\it 5}^{\rm w} = \frac12 (\tilde {\bf F}_{\it 5}^{\rm w} + \tilde {\bf F}_{\it 5}^{\rm wo}) =  \frac12 \tilde {\bf F}_{\it 5},
\ee
with the indices now runing over the full ten-dimensional coordinates including $y'$ and the overall factor $\frac12$. 

Finally, we decompose ten-dimensional gamma matrices into nine-dimensional ones. From the charge conjugation property of twelve-dimensional spinors, introduced in the appendix, 
$$
  \Gamma^{y'} \Psi_{\rm Maj} = \Gamma \Psi_{\rm Maj} , \quad \Gamma^{y'} \Psi_{\rm L} = \Gamma \Psi_{\rm R}, \quad \quad \Gamma^{y'} \Psi_{\rm R} = \Gamma \Psi_{\rm L}.
$$
We see that the consistent structure is
\be
 \upgamma^{y'} \equiv \upgamma^y = i \begin{pmatrix} 0 & - {\bf 1} \\ {\bf 1} & 0 \end{pmatrix},
\ee
as in (\ref{gammay}).
Therefore the only possible choice for the remaining matrices is \cite{Bergshoeff:1995as}
\be \label{9Dgamma}
 \upgamma^\alpha = \gamma^\alpha \sigma^1, \quad \alpha=0,1,2,\dots,8,
\ee
with the tensor product understood, and $\gamma^8 = \prod_{\alpha=0}^7 \gamma^\alpha, \sigma^1 = \left(\begin{smallmatrix} 0 &1 \\ 1 & 0 \end{smallmatrix}\right)$.
Thus we have transformations of IIB supergravity to the leading order \cite{Becker:2007zj,Bergshoeff:1999bx}
\begin{align}
\delta\begin{pmatrix} 0 \\  \lambda \end{pmatrix}
   &= -  \frac{i}{2\tau_2}\gamma^{\upmu}  \partial_{\upmu} \tau \sigma^1  \begin{pmatrix} \ve \\ 0 \end{pmatrix}
 + \frac{i}{4\tau_2} \left(\tilde {\bf F}_{\it 3} +i \tau_2 {\bf H}_{\it 3} \right)\sigma^1  \begin{pmatrix} \ve^* \\ 0 \end{pmatrix}, \\
 \delta\begin{pmatrix} \psi_{\upmu} \\ 0 \end{pmatrix}  &= \left(\nabla_{\upmu} + \frac{i}{8 \tau_2}  {\bf F}_{\1} \gamma_{\upmu} + \frac{i}{16 \tau_2} \tilde {\bf F}_{\it 5} \gamma_{\upmu} \right)  \begin{pmatrix} \ve \\ 0 \end{pmatrix}
   + \frac18 \left( \gamma_{\upmu} {\bf H}_{\3} - {\bf H}_{\3} \gamma_{\upmu}- \frac{i}{\tau_2} \tilde {\bf F}_{\it 3} \Gamma_{\upmu} \right)  \begin{pmatrix} \ve^* \\ 0 \end{pmatrix}.
 \end{align}
Here the complex structure $\tau$ of the torus defined in (\ref{cs}) reappears as axion-dilaton $\tau = A + i e^{-\phi}$ with the IIB dilaton $e^{-\phi}$.
 
Observe that the dilatino should have the opposite chirality to the supersymmetry generator, by $\sigma^1$ dependence from the ten-dimensional embedding of the gamma matrix (\ref{9Dgamma}). On the other hand, gravitinos have the same chirality, because we have always even number of gamma matrices resulting even multiple of $\sigma^1$. The chirality is hidden in the gamma matrix structure, since each gamma matrix flips the ten-dimensional chirality. 

Summarizing, without expecting any low-energy theory, the choice (\ref{12DWeyl}) gave a particular presentation of nine-dimensional supergravity in the above form. Decompactification associating to the $y'$ direction gave rise to a chiral theory of IIB supergravity in ten dimension.

This is counterintuitive to the known fact that compactification of a chiral theory on a torus gives rise to parity symmetric theory. For instance compactificaiotn of four-dimensional ${\cal N}=1$ chiral theory on a torus gives rise to two dimensional $(2,2)$ theory. As argued \cite{Schwarz:1996bh}, nine-dimensional supergravity is chiral, with massive modes which is absent in this four dimensional analogy. The massive fields are provided by KK modes along $y'$-direction (\ref{KKtower}). In the M-theory limit, this is understood as wrapped modes of M2-branes. In terms of type IIB language this is winding mode in the dual IIA string. The little algebra for ten-dimensional chiral massless fields is the same as that of nine-dimensional massive fields.

\section{Reduction to IIA supergravity}

Our main focus in this paper is on IIB supergravity, for which the notations are prepared. Since the twelfth dimension concretely realizes the $T$-duality as compactification, it would be useful to compare the geometry and field contents of the IIA and IIB theory.

\subsection{Field reduction} \label{s:reduction}

We go back to the nine-dimensional theory and consider the decompactification of the $y$ direction, which is to be understood as type IIA supergravity. Indeed, it is understood as compactification of eleven-dimensional supergravity on the circle along the $x$-direction, which is again understood as twelve-dimensional supergravity compactified on the circle along the $y'$-direction
\be
 ds^2 = L^{-1} G_{\upmu \upnu} dx^{\upmu} dx^{\upnu} + L^2 (dx + A_{\upmu} dx^{\upmu})^2 + r^2 dy^{\prime 2}, \quad \upmu,\upnu = 0,1,\dots,9,
\ee
where we call this coordinate $x$,
 gives IIA supergravity. Here the ten dimensional vector $A_{\upmu} = \{a_\mu, \tau_1 \} $ has $\tau_1$ as a component in the $y$-direction, as shown in Table \ref{t:fields}. We define
$$ G_{\upmu \upnu \uprho \upsigma} = r^{-1} \G_{\upmu \upnu \uprho \upsigma y'} \equiv  F_{\upmu \upnu \uprho \upsigma}, 
\quad G_{\upmu \upnu \uprho x} = r^{-1} \G_{\upmu \upnu \uprho x y'} \equiv - H_{\upmu \upnu \uprho  }. $$
Then
\begin{align}
  \G_{\upalpha \upbeta \upgamma \updelta y'} 
   &= E^{\upmu}_{\upalpha} E^{\upnu}_{\upbeta} E^{\uprho}_{\upgamma} E^{\upsigma}_{\updelta} E^{y'}_{y'} \G_{\upmu \upnu \uprho \upsigma y'} 
   + E^{\upmu}_{\upalpha} E^{\upnu}_{\upbeta} E^{\uprho}_{\upgamma} E^{x}_{\updelta } E^{y'}_{y'} \G_{\upmu \upnu \uprho xy'} \nn \\
&=  E^{\upmu}_{\upalpha} E^{\upnu}_{\upbeta} E^{\uprho}_{\upgamma} E^{\upsigma}_{\updelta}  G_{\upmu \upnu \uprho \upsigma} + E^{\upmu}_{\upalpha} E^{\upnu}_{\upbeta} E^{\uprho}_{\upgamma} E^{x}_{\updelta} \G_{\upmu \upnu \uprho x} \label{IIAF4} \\
 &=  L^2(F_{\upalpha \upbeta \upgamma \updelta} - 4 H_{[\upalpha \upbeta \upgamma} A_{\updelta]}), \nn
\end{align}
Since the zw\"olfbein $E$ has only diagonal component in the $y'$ direction, we can always convert the antisymmetric tensor fields of twelve and eleven dimensions.
Note that we are using the new metric. Further compactification on a circle in the $y$-direction, for which now we can use the metric (\ref{themetric}), 
gives
\begin{align*}
\G_{\alpha \beta \gamma\delta y'} & = L^2(F_{\alpha \beta \gamma \delta} - 4 a_{ [\alpha} H_{\beta \gamma \delta]} + 4 b_{ [\alpha} F_{\beta \gamma \delta]y} 
+ 12 a_{[\alpha} b_\beta H_{\gamma \delta]y} ), \\
 \G_{\alpha \beta \gamma y  y'} & =  L^{1/2} \tau_2^{-1} ( F_{[\alpha \beta \gamma]y} - a_y H_{\alpha \beta \gamma} 
 +3 a_{[\alpha} H_{\beta \gamma]y}- 3 a_y b_{[\alpha} H_{\beta \gamma]y} ) 
\end{align*}
This agrees well with the computations (\ref{G5expA}) and (\ref{ExptoF3}) in the appendix, with a little change of notations.
The fields having an index on $y$ follows the rule (\ref{indexrule})
$$ L^{1/2} \tau_2^{-1} F_{\alpha \beta \gamma y} = R_{y}   L^{1/2} \tau_2^{-1} F_{\alpha \beta \gamma y}^{(10)} = L^2 F_{\alpha \beta \gamma y}^{(10)} . $$ 
All of these reproduce the rules of the dimensional reduction of eleven-dimensional supergravity to type IIA supergravity.

In the fermionic sector, we follow the conventional route. First take the Majorana generator (\ref{SUSYMajorana}) and take the first two entries which are eleven- and ten-dimensional Majorana spinor. Now the ten-dimensional gravitino have a combination from $\Psi_x$, so that
\be
\psi_\alpha^{\rm new} \leftarrow L^{-1/4} \left(  \Psi_\alpha + \frac12 \Gamma_\alpha \Gamma^x \Psi_x  \right), \quad 
\lambda^{\rm new} \leftarrow L^{-1/4}  \Psi_x , \quad
\begin{pmatrix} \epsilon_1 \\ \epsilon_2 \end{pmatrix}^{\rm new} \leftarrow L^{-1/4}\begin{pmatrix} \epsilon_1 \\ \epsilon_2 \end{pmatrix}
\ee
The dilatino is simply defined as the component $\Psi_x$ without an extra gamma factor. Therefore there is no chirality flip.

\subsection{Generalized $T$-duality and decompactification limits}

We summarize the relations among supergravity theories, controlled by the shape and the size of the torus.
From the requirement of ten-dimensional Lorentz invariance of type IIB supergravity, we obtained the relation (\ref{r}) among the radii of the circles in the $x,y,y'$ directions.
The final metric after the rescaling (\ref{rescaling}) is
\begin{equation} \label{thenewmetric} 
 \begin{split}
 \d s^2 = &\ L^2 \left(\d x +  \tau_1 \d y +(a_{\mu} - \tau_1 b_{ \mu}) \d x^\mu  \right)^2 
                    + L^2 \tau_2^2 \left( \d y -  b_{\mu } \d x^\mu\right)^2 \\
               &    + L^{-4} \tau_2^{-2} \d y^{\prime 2} +L^{-1} g_{\mu \nu} dx^\mu dx^\nu.
 \end{split}
\end{equation}
We have considered the following limits.\begin{itemize}
\item Ten-dimensional IIA supergravity: $L \to 0, L\tau_2 \to \infty$. The complex structure $\tau_2$ should grow faster than $L^{-1}$, which condition is achieved by $L\to 0,r\to 0$. We have ten-dimensional metric
$$ ds_{\rm IIA}^2 = L^3 \tau_2^2 dy^2 + g_{\mu \nu} dx^\mu dx^\nu. $$
We identify IIA dilaton and string coupling as $g_{\rm IIA} =L^{3/2}$.
\item 
Eleven-dimensional supergravity: $L \to \infty$. With finite $\tau_2$, the torus decompactifies and we have $r \to 0$.
Using the identification of IIA coupling, the radius of the $x$-circle is $L \ell = g_{\rm IIA}^{2/3}\ell = g_{\rm IIA} \ell_{\rm s}$ and the decompactification to eleven-dimensional supergravity is done in the strong coupling limit $g_{\rm IIA} \to \infty$.

\item Ten-dinemsional IIB supergravity: $L \to 0$. Keeping $\tau_2$ finite sends $r\to \infty$ and shrinks the torus, whose area is $L^2 \ell^2 \tau_2$. The ten-dimensional metric is
$$ ds_{\rm IIB}^2 = \frac{1}{L^3 \tau_2^2} dy^{\prime 2} + g_{\mu \nu} dx^\mu dx^\nu. $$
Although the extra torus is shrunk, there is axion-dilaton field in the spectrum, so that we are left with the complex structure as a footprint. The K\"ahler modulus $L$ is absent in this limit.
\end{itemize}
These limit shows the $T$-duality relation between IIA and IIB theories, exchanging the $y$- and $y'$- directions. We thus have obtained and identified the fields, using the IIA radius $L_A = L^{3/2} \tau_2$,
$$ g_{\rm IIB} = \frac{1}{L_A} g_{\rm IIA} 
= \tau_2^{-1}, \quad  A = \tau_1$$
Since we have obtained $T$-duality as different dimensional reduction, we have generalized $T$-duality also as exchange. 
The Kaluza--Klein momenta (\ref{KKmasses}) along the $y'$-direction becomes 
\be
 M_{k}^2 = \frac{k^2}{\ell^2 \langle r \rangle^2} = \frac{ k^2 \langle L\rangle^4 \tau_2^2} {\ell^2} .
\ee
winding of M2-brane along the torus of the area $\langle L\rangle^2  \ell^2 \tau_2$.

\section{Discussion}
We have constructed twelve-dimensional supergravity action, whose compactification yields eleven-dimensional and type IIB supergravity actions. We also obtained supersymmetry transformations at linear order of fermions, which agree with the known rules of IIB supergravity. Of course, further compactifications shall yield type IIA and type I supergravity actions, too. 

The key observation is, that we have always have at least one compact dimension. First we demanded the twelfth dimension to be a compact circle and be orthogonal to the other directions. Although we have introduced the graviton $e_{y'}^{y'} \equiv r (x^m,y')$ in the new direction, requiring a decompactification forced us to identify this with the components of other directions. Therefore we have we have the same degrees of freedom those of eleven-dimensional supergravity. This is how we circumvent the no-go theorem by Nahm.

In showing this, compactification of two dimensions and decompactification of one dimension is crucial. There is no direct way to obtain covariant IIB supergravity from twelve dimensions. The off-diagonal components of the graviton of IIB theory emerges from components of antisymmetric tensor fields in M-theory $E^\mu_{y'} = C_{\mu x y}$. This is well-known problem showing the notion of dimension and fundamental degrees of freedom is subtle. Note that the metric tensor is not an observable and in the low energy we cannot its interaction from the interaction of the antisymmetric tensor field. One safe way to understand this is to start with three dimensions $x,y,y'$ compact and to consider various decompactification limit. Each vacua giving IIA and IIB has different Poincar\'e symmetry are controlled by the the moduli of the torus $L$ and $\tau$. However the full decompacitfication limits should be dealt with care. 

The following summary gives us some insights on the meaning of the twelfth dimension.

The four-form field of IIB supergravity is naturally understood as dimensional reduction of twelve-dimensional four-form field. Its tensor structure is naturally obtained by lifting three-form field $C_{\it 3}$ of eleven-dimensional supergravity to twelve-dimensional $\C_{\it 4}$. It follows that an M2-brane is understood as a 3-brane wrapped on the circle $S_{y'}$.

Half of the components of four-form field of IIB supergravity is {\em defined} by what is known as self-duality condition in this framework. This condition is imposed to the combination $\tilde F_{\it 5}$ in our notation, not $F_{\it 5}$. It originates from Hodge duality in twelve-dimension, which imposes the correct index structure. 

Spinorial representation in twelve-dimension shows us how the IIA non-chiral fermion is re-arranged to IIB ones. It comes from a special property in $4n$-dimension in which Majorana spinor is converted to Weyl. However we have just embedded the eleven-dimensional spinors in the twelve-dimension while keeping only the eleven-dimensional Lorentz symmetry.

We can concretely perform $T$-duality as dimensional reduction, as shown in Figure \ref{f:relations}, and its generalization can be done to the objects of eleven-dimensional supergravity. Although we have known what rules are necessary to convert between the fields of IIA and IIB supergravities, and could exchange them by hand, this compactification tells us how to perform $T$-duality. Relation between $T$-dual radii are consequence of Lorentz symmetry.

To actually perform $T$-duality from IIA to IIB, we need tower of KK states. They are provided by compactification of the dynamical $y'$-direction to recover the IIB string theory in ten dimension.
The wrapping modes of M2-brane is now captured as KK fields: the wrapping number becomes the KK momentum in the $y'$-direction. In other words, we have a way to deal with winding or wrapping of classical strings and membranes in the effective field theory, which enables us to capture physics at the self-dual radius.

Although the theory has only complex structure for the extra torus than ten dimensions in the IIB spacetime, we can make use of this information. For example, obtaining realistic string constructed model in four dimensions can directly make use of eight extra dimensions.

Lift of three-form field of eleven dimensional supergravity to twelve-dimensional four-form field predicts three-brane solution in the twelve-dimensional context.

In this way we can discover strings and membranes. After decompactification of the twelfth dimension, the ten-dimensional Poincar\'e symmetry become nontrivial. Imposing Lorentz covariance to four-form field gives us a nontrivial condition determining the size of the twelfth dimension. This should be, because $T$-duality relates the radii of dual circles.
Also reproduction IIB supergravity by dimensional reduction of twelve-dimensional supergravity, without extra degrees of freedom sets the string frame, reveals the string length in terms of a fundamental length. With the relation of radii, we can discover a classical string solution with the tension.

We hope this effective action can take into account the bulk and gravity action and facilitates direct calculation of realistic four-dimensional action obtained from F-theory compactification.

\subsection*{Acknowledgements} 
The author is grateful to Lara Anderson, James Gray, Jihn E. Kim, Nakwoo Kim, Jeong-Hyuck Park, and Soo-Jong Rey for discussions. Especially he thanks to Imtak Jeon for the help on technical details. This work is partly supported by the National Research Foundation of Korea with grant number 2012-R1A1A1040695.

\appendix

\section{Spinor in twelve dimension}

We consider twelve-dimensional spinor with Lorentz signature $(1,11)$. The Dirac spinor, transforming under the Clifford algebra, has 32 complex components. The minimal spinor here has 32 real components. It can be either Majorana, satisfying the further condition
\be \label{12DMajorana}
 \Psi_{\rm Maj} = B^* \Psi_{\rm Maj}^*
\ee
with the charge conjugation matrix 
\be
 B =  \Gamma \Gamma^{y'}  = i \begin{pmatrix} & & -{\bf 1} &   \\ & &  & {\bf 1}  \\ -{\bf 1} &   & & \\  & {\bf 1} & & \\ \end{pmatrix},\quad B^*B = 1,
\ee
or Weyl, the eigenstate of chirality operator $\Gamma$
\be \label{12DWeyl}
 \Gamma \Psi_{\rm L} = -\Psi_{\rm L}, \text{ or } \Gamma \Psi_{\rm R} = \Psi_{\rm R}.
\ee
These spinors are all exchangeable, as a special property in $4n$-dimension,
$$
\Psi_{\rm Maj} = \begin{pmatrix}  \psi_1 \\ \psi_2 \\ i \psi_1^* \\ -i \psi_2^* \end{pmatrix}  \Longleftrightarrow
\Psi_{\rm L} = \begin{pmatrix} \psi_1 \\ 0 \\ 0 \\ -i  \psi_2^* \end{pmatrix} \Longleftrightarrow
\Psi_{\rm R} =  \begin{pmatrix} 0 \\  \psi_2 \\ i  \psi_1^* \\ 0 \end{pmatrix}.
$$
We have the following relataion
$$
  \Gamma^{y'} \Psi_{\rm Maj} = \Gamma \Psi_{\rm Maj}^* , \quad \Gamma^{y'} \Psi_{\rm L} = \Gamma \Psi_{\rm R}^*, \quad \quad \Gamma^{y'} \Psi_{\rm R} = \Gamma \Psi_{\rm L}^*.
$$
Weyl spinors of opposite chirality are complex conjugate 
\be
 \Psi_{\rm Maj} =  \begin{pmatrix}  \psi_1 \\ \psi_2 \\ 0 \\0 \end{pmatrix} + B^* \begin{pmatrix}  \psi_1^* \\  \psi_2^* \\ 0\\ 0 \end{pmatrix}  = \Psi_{\rm L} + B^* \Psi_{\rm L}^* = \Psi_{\rm R} + B^* \Psi_{\rm R}^*.
\ee

With Lorentz signature $(1,10)$ we may have minimal spinor with 32 real components, which is however not of our concern. It is a spacelike torus on which F-theory is compactified to yield type IIB string theory with varying axion-dilator with the Lorentz signature $(1,9)$.

\section{Formulae}

Here we summarize the formulae appeared in the main text and supplement details of derivations. We start from nine dimension plus three compact dimensions.
We use the metric in the string frame (\ref{thenewmetric}), which obtained from the metric (\ref{themetric}) after the rescaling  (\ref{rescaling}).
The corresponding inverse-zw\"olfbein is
\begin{equation}
 E^M_A = 
\begin{pmatrix}
 L^{1/2} E^\mu_{\alpha}  & 0 & 0 &0 \\ 
 L^{1/2} b_{\alpha} & L^{-1} \tau_2^{-1} & 0 &0  \\
 -L^{1/2} a_{\alpha} & -\tau_1 \tau_2^{-1} L^{-1} & L^{-1} & 0 \\   
 0 & 0 & 0 & L^{3/2} \tau_2^2 
\end{pmatrix},
\quad e_M^A = (E^{-1})^A_M,\quad G_{MN} = e_M^A e_N^B \eta_{AB}.
\end{equation}
We have tensors in components in local Lorentz frame as in (\ref{fieldrecuction})
$$ \G_{abcdy'} = \G_{mnpqy'} E^m_a E^n_b E^p_c E^q_d E^{y'}_{y'} = G_{mnpq} E^m_a E^n_b E^p_c E^q_d  $$
where we used the definition.

\begin{align}
 \C_{\alpha \beta \gamma y'} &= L^{3/2} (  A_{[\alpha \beta \gamma]y'} - 3 a_{[\alpha} B_{\beta \gamma]} + 3 b_{[\alpha} A_{\beta \gamma]}  
  - 6 a_{[\alpha} b_{\beta} K_{\gamma]}),  \\
 \C_{\alpha \beta x y'} &=  B_{\alpha \beta} + 2b_{[\alpha} K_{\beta]}, \\
 \C_{\alpha \beta y y'} &= \tau_2^{-1}  (A_{\alpha \beta} - \tau_1 B_{\alpha \beta} + 2a_{[\alpha} K_{\beta]} 
 - 2 \tau_1 b_{[\alpha} K_{\beta]} ),\\
 \C_{\alpha x y y'} &= L^{-3/2}\tau_2^{-1} K_\alpha,\\
 C_{\alpha \beta \gamma \delta  xy} &=  (r^{2}L^{2} \tau_2 )^{-1} A_{\alpha \beta \gamma \delta} .
 \label{C6reduction} 
 \end{align}
 Here we have reflected the $y'$ metric component, but these can be regarded as eleven-dimensional tensors as well, by simply dropping $y'$ index.
 And we have the corresponding field strengths
 \begin{align}
 \G_{\alpha \beta \gamma\delta y'} & =L^{2}( F_{[\alpha \beta \gamma \delta] y'} - 4 a_{ [\alpha} H_{\beta \gamma \delta]} + 4 b_{ [\alpha} F_{\beta \gamma \delta]} 
+ 12 a_{[\alpha} b_\beta H_{\gamma \delta]})   \nn \\
& = L^{2} ( F_{[\alpha \beta \gamma \delta] y'} + 4 A_{y' [\alpha} H_{\beta \gamma \delta]} -4 B_{y' [\alpha} F_{\beta \gamma \delta]} \nn \\
&\quad+ 12A_{y'[\alpha} K_{\beta} H_{\gamma \delta] y'} -12 B_{y' [\alpha} K_{\beta} F_{\gamma \delta]y'}) \label{G5expA}  \\
& = L^{1/2} \tau_2^{-1} (F^{(10)}_{[\alpha \beta \gamma \delta] y'} 
+ 2A^{(10)}_{y' [\alpha} H^{(10)}_{\beta \gamma \delta]} 
 - 3 A^{(10)}_{[ \alpha \beta} H^{(10)}_{\gamma \delta]y'}
 - 2 B^{(10)}_{y' [\alpha} F^{(10)}_{\beta \gamma \delta]} 
  +3 B^{(10)}_{[ \alpha  \beta}  F^{(10)}_{ \gamma \delta]y'} )\nn \\
&=L^{1/2} \tau_2^{-1}  \left(F^{\rm w (10)}_{\it 5}- \frac12 A_{\2}^{(10)}  \wedge H_{\3}^{(10)} + \frac12 B_{\2}^{(10)}  \wedge F_{\3}^{(10)} \right)_{[\alpha \beta \gamma \delta y']} \nn \\
 \G_{\alpha \beta \gamma x  y'} & = L^{1/2}(  H_{\alpha \beta \gamma} +3 b_{[\alpha} H_{\beta \gamma]}) \nn \\
  &= L^{1/2}( H_{\alpha \beta \gamma} +3 K_{[\alpha} H_{\beta \gamma]} )  \label{H3A} \\
  &= L^{1/2} H^{(10)}_{\alpha  \beta \gamma} , \nn \\
 \G_{\alpha \beta \gamma y  y'} & =L^{1/2}   \tau_2^{-1} ( F_{\alpha \beta \gamma} - \tau_1 H_{\alpha \beta \gamma} 
 +3 a_{[\alpha} H_{\beta \gamma]}- 3 \tau_1 b_{[\alpha} H_{\beta \gamma]} )  \nn \\
 &=L^{1/2}  \tau_2^{-1} ( \tilde F_{\alpha \beta \gamma} + 3 \tau_1 K_{[\alpha} \tilde F_{\beta \gamma]y'}     )  \label{ExptoF3} \\
 &= L^{1/2} \tau_2^{-1} \tilde F^{(10)}_{ \alpha \beta \gamma},  \nn\\
  \G_{\alpha \beta x y y'} &=  ( L \tau_2)^{-1} H_{\alpha \beta} , \label{Kkinetic} \\
  & =( L \tau_2)^{-1} H^{(10)}_{\alpha \beta}  
   \end{align}
Here $\tilde F^{(10)}_{ \alpha \beta \gamma} = F^{(10)}_{ \alpha \beta \gamma} - A H^{(10)}_{ \alpha \beta \gamma}.$
Also   
\be
  (dC)_{\alpha \beta \gamma \delta \epsilon xy} = (r^{2}L^{2} \tau_2 )^{-1} F_{\alpha \beta \gamma \delta \epsilon} 
 \label{G7reduction}.
\ee

Since $F_{\alpha \beta \gamma \delta y'}$ itself does not have reduce into several pieces, there is no associated decompactification term with $K_{\1}$. 
Also note that although in the terms involving $K_{\1}$, $y'$ appears twice but this is not contradictory to the total antisymmetric index structure.
The combination (\ref{G5expA}), where the $y$-component is missing, and (\ref{ExptoF3}) with the $y$-compoent, can also give the IIA field (\ref{IIAF4}), after using the decompactification field $b_\mu$.

The following `dual fields' do not appear in the action.
\begin{align}
 C_{\alpha \beta \gamma \delta xy} &= L^2 \tau_2 A_{\alpha \beta \gamma \delta} \\
   (dC)_{\alpha \beta \gamma \delta \epsilon xy} &= L^2 \tau_2 F_{\alpha \beta \gamma \delta \epsilon} 
 \label{G7reduction}.
 \end{align}
 The other components of $C_6$ come from the dual field (\ref{C4fromC6})
\begin{equation} \label{C3G4} 
 C_{\it 3} \wedge G_{\it 4}|_{xy} = L^2 \tau_2 ( A_{\it 2} \wedge H_{\it 3} -B_{\it 2 } \wedge F_{\it 3} 
 - A_{\it 3} \wedge H_{\it 2}-B_{\it 1} \wedge F_{\it 4} ).
\end{equation}
where $|_{xy}$ means we have set two of the components to be $x$ and $y$, but in $C_{\it 1} \wedge G_{\it 4}$ we have no that restriction. Therefore $(dC_6 - \frac12 C_{\it 3} \wedge G_{\it 4})|_{xy}$ provides the rest of the tensor (\ref{G5exp10D}), without the $y'$ component. This is in agreement with the self-duality condition for $\tilde F_5$ as well.

Some definitions 
\begin{align}
{\mathbb F}_{\1} &\equiv \gamma^{\alpha} \partial_\alpha A \label{F1}\\
\tilde {\mathbb F}_{\it 5}^{\rm w} &\equiv  \frac{1}{5!} \tilde F_{\alpha \beta \gamma \delta y'} \gamma^{\alpha \beta \gamma \delta y'} =  \frac{1}{4!} \tilde F_{\alpha \beta \gamma \delta y'} \gamma^{\alpha \beta \gamma \delta }\gamma^{y'}    , \\
 \tilde {\mathbb F}_{\3} & \equiv \frac{1}{3!} \tilde F_{\alpha \beta \gamma } \gamma^{\alpha \beta \gamma }, \\
 \tilde {\mathbb H}_{\3} & \equiv \frac{1}{3!} (H_{\alpha \beta \gamma } + 3 b_{[\alpha}  H_{\beta \gamma]}) \gamma^{\alpha \beta \gamma }, \\
 {\mathbb H}_{\2} & \equiv \frac{1}{2} H_{\alpha \beta} \gamma^{\alpha \beta}.
\end{align}
A two-form in IIA side can be viewed as three-form in IIB side
\begin{align}
  \frac{1}{2} {\bf h}_2 & \equiv  \partial_{\alpha} h_\beta  \gamma^{\beta \gamma} \equiv \frac{1}{2} h_{\beta \gamma }  \gamma^{\beta \gamma} (\gamma^{y'})^2  =  \frac{1}{3!} h_{\beta \gamma y'}  \gamma^{\beta \gamma y'} \Gamma^{y'} = {\mathbb H}_{\3}^{\rm w} \gamma^{y'}, \\
\tilde{\mathbb F}_{\3}^{\rm w} & \equiv \frac{1}{3!} (f_{ \beta\gamma y'} - A h_{\beta \gamma y'})\gamma^{ \beta\gamma y'} = \frac{1}{2}(f_{\alpha \beta} - A h_{\alpha \beta}) \gamma^{\alpha \beta} \gamma^{y'}. \label{F3w}
\end{align} 
All the gamma matrices here are nine-dimensional. 

The eleven-dimensional supersymmetry generators embedded in twelve dimension are reduced in nine dimension as follows
 \be \label{Gammaaction}
 \Gamma^x {\cal E} \to   \varepsilon^* , \quad \Gamma^y {\cal E} \to   - i \varepsilon^* , \quad \Gamma^x \Gamma^y {\cal E}\to    - i \varepsilon^*, \quad \Gamma^x \Gamma^{y'} {\cal E} \to i \ve, \quad \Gamma^y \Gamma^{y'} {\cal E} \to \ve.
\ee

Some useful relations:
\be  \label{identity}
ab \nabla^2 (a b)^{-1} = 2 a^{-2} (\partial_\mu a)^2 + 2 b^{-2} (\partial_\mu b)^2 - a^{-1} \nabla^2 a - b^{-1} \nabla b. 
\ee
In the main text we consider the case $a=L,b= L \tau_1$. We also have $$ F_{ab} \Gamma^b = \frac{1}{2} (\Gamma_a {\mathbb F}_{\2} - {\mathbb F}_{\2} \Gamma_a). $$

\end{document}